\documentclass[letter,bibyear]{aa} 
\usepackage{epsfig}
\usepackage{amsmath}
\usepackage{subfigure}
\usepackage{natbib}
\usepackage{multirow}
\usepackage{color}
\usepackage{longtable}
\usepackage{graphicx}
\usepackage{epstopdf}
\usepackage{booktabs}
\usepackage{txfonts}

\newcommand{\jmst}{J.~Mol.~Struct.}   

\begin{document}

\title{Discovery of the acetyl cation, CH$_3$CO$^+$, in space and in the laboratory
\thanks{Based on observations carried out
with the Yebes 40m telescope (projects 19A003,
20A014, and 20D15) and the Institut de Radioastronomie Millim\'etrique (IRAM) 30m telescope. The 40m
radiotelescope at Yebes Observatory is operated by the Spanish Geographic Institute
(IGN, Ministerio de Transportes, Movilidad y Agenda Urbana). IRAM is supported by INSU/CNRS
(France), MPG (Germany), and IGN (Spain).}}

\author{
J.~Cernicharo\inst{1},
C.~Cabezas\inst{1},
S.~Bailleux\inst{2},
L.~Margul\`es\inst{2},
R.~Motiyenko\inst{2},
L.~Zou\inst{2},
Y.~Endo\inst{3},
C.~Berm\'udez\inst{1},
M.~Ag\'undez\inst{1},
N.~Marcelino\inst{1},
B.~Lefloch\inst{4},
B.~Tercero\inst{5,6}, and
P.~de Vicente\inst{5}
}

\institute{Grupo de Astrof\'isica Molecular, Instituto de F\'isica Fundamental (IFF-CSIC), C/ Serrano 121, 28006 Madrid, Spain.
\email jose.cernicharo@csic.es
\and
Univ. Lille, CNRS, UMR 8523 - PhLAM - Physique des Lasers Atomes et Mol\'ecules, 59000 Lille, France
\and Department of Applied Chemistry, Science Building II, National Chiao Tung University, 1001 Ta-Hsueh Rd., Hsinchu 30010, Taiwan
\and
CNRS, IPAG, Univ. Grenoble Alpes, F-38000 Grenoble, France
\and Observatorio Astron\'omico Nacional (IGN), C/ Alfonso XII, 3, 28014, Madrid, Spain.
\and Centro de Desarrollos Tecnol\'ogicos, Observatorio de Yebes (IGN), 19141 Yebes, Guadalajara, Spain.
}

\date{Received; accepted}

\abstract{Using the Yebes 40m and IRAM 30m radiotelescopes, we detected two series
of harmonically related lines in space that can be fitted to a symmetric rotor.
The lines
have been seen towards the cold dense cores TMC-1, L483, L1527, and L1544.
High level of theory \emph{ab initio} calculations indicate that the best possible
candidate is the acetyl cation, CH$_3$CO$^+$, which is the most stable product
resulting from the protonation of ketene. We have produced
this species in the laboratory and observed its rotational transitions $J_u$ = 10 up to $J_u$ = 27.
Hence, we report the discovery of CH$_3$CO$^+$ in space based on our observations, theoretical
calculations,
and laboratory experiments. The derived rotational and distortion constants allow us to
predict the spectrum of CH$_3$CO$^+$ with high accuracy up to 500 GHz.
We derive an abundance ratio
$N$(H$_2$CCO)/$N$(CH$_3$CO$^+$)$\sim$44. The high abundance of the protonated form of H$_2$CCO is
due to the high proton affinity of the neutral species. The other isomer, H$_2$CCOH$^+$, is found
to be 178.9 kJ mol$^{-1}$ above CH$_3$CO$^+$. The observed intensity ratio between the $K$=0 and $K$=1
lines, $\sim$2.2, strongly suggests that the $A$ and $E$ symmetry states have suffered
interconversion processes due to collisions with H and/or H$_2$, or during their formation
through the reaction of H$_3^+$ with H$_2$CCO.
}


\titlerunning{Discovery of CH$_3$CO$^+$ in space}
\authorrunning{Cernicharo et al.}

\maketitle

\section{Introduction}
The cold dark core TMC-1 presents an interesting chemistry. It produces a significant
number of the molecules detected in space, in particular
long neutral carbon-chain radicals and their anions (see e.g. \citealt{Cernicharo2020a, Marcelino2020a},
and references therein) as well as cyanopolyynes (see \citealt{Cernicharo2020b} and \citealt{Xue2020}, and
references therein). The presence in this object of O-bearing carbon chains, such as
C$_2$O \citep{Ohishi1991}, C$_3$O \citep{Matthews1984},
HC$_5$O \citep{McGuire2017}, HC$_7$O \citep{Cordiner2017}, HCCO, and HC$_3$O$^+$
\citep{Cernicharo2020c}, is a surprising result that has not
yet been fully accounted for by 
chemical models.

The abundance of polyatomic cations in cold interstellar clouds is relatively low because they react
fast with electrons.
Interestingly, all polyatomic cations detected in cold clouds are protonated forms of stable and
abundant molecules. Chemical models and observations suggest a trend in which the protonated-to-neutral
abundance ratio [MH$^+$]/[M] increases with the proton affinity of
M \citep{Agundez2015,Cernicharo2020c,Cernicharo2020d,Marcelino2020a}.

It has been suggested that some O-bearing cations are sufficiently long-lived to be
abundant \citep{Petrie1993}. We have recently reported the discovery of the cation HC$_3$O$^+$ in TMC-1
\citep{Cernicharo2020c}.
In this letter, we report the detection of two series of lines that are harmonically related towards the
cold dark core TMC-1. These lines can be fitted as the $K$=0 and $K$=1 lines of
a symmetric rotor. From the astronomical data and the derived rotational constants, together with
high-level \emph{ab initio} calculations, we suggest CH$_3$CO$^+$ as the best possible carrier.
We have performed microwave laboratory experiments that fully support this hypothesis: We detected 79 rotational transitions near the predicted frequencies from the astronomical
constants. Hence, we report the discovery in space and in the laboratory of CH$_3$CO$^+$ (acetyl cation),
which is the most stable isomer resulting from the protonation of ketene (H$_2$CCO).
The presence of CH$_3$CO$^+$ can be expected on the basis of the high abundance of
H$_2$CCO in TMC-1 and its large proton affinity (825.3 kJ mol$^{-1}$; \citealt{Traeger1982}).
An anomalous abundance ratio of 2.2 is found between the $A$ and $E$ symmetry species of CH$_3$CO$^+$.
We discuss these results in the context of state-of-the-art chemical models and in terms
of the interconversion of $E$-CH$_3$CO$^+$ into $A$-CH$_3$CO$^+$ through the formation process
of the molecule or by collisions with H and/or H$_2$.

\section{Observations}
New receivers, built
as part of the Nanocosmos
project\footnote{\texttt{https://nanocosmos.iff.csic.es/}} and installed at the Yebes 40m radio telescope, were used
for the observations of TMC-1. The Q-band receiver consists of two high electron mobility transistor (HEMT) cold amplifiers 
that cover the
31.0-50.3 GHz band with horizontal and vertical polarizations. Receiver temperatures
vary from 22 K at 32 GHz to 42 K at 50 GHz. The spectrometers are $2\times8\times2.5$ GHz fast Fourier transform 
spectrometers (FFTs)
with a spectral resolution
of 38.1 kHz, providing the whole coverage of the Q-band in both polarizations.
The main beam efficiency varies from 0.6 at 32 GHz to 0.43 at 50 GHz \citep{Tercero2020}.

The observations that led to the line survey in the Q-band towards TMC-1
($\alpha_{J2000}=4^{\rm h} 41^{\rm  m} 41.9^{\rm s}$ and $\delta_{J2000}=+25^\circ 41' 27.0''$)
were performed in several sessions,
between November 2019 and February 2020. The observing procedure was
frequency switching with a frequency throw of 10\,MHz. The nominal spectral
resolution of 38.1 kHz was used for the final spectra.
In these runs, two
different frequency coverages were observed, 31.08-49.52 GHz and 31.98-50.42 GHz. This permits the user to check that no
spurious ghosts are produced in the down-conversion chain, in which the signal coming from the receiver is
down-converted to 1-19.5 GHz and then split into eight bands with a coverage of 2.5 GHz, each of which
are analysed
by the FFTs. Additional data were taken in October 2020 to improve the line survey at some frequencies and to
further check the consistency of all observed spectral features. These observations were also performed
in frequency switching but with a throw of 8 MHz.
The sensitivity varies along the
Q-band between 0.5 and 2.5 mK, which is a considerable improvement compared to previous line surveys in the 31-50 GHz frequency range
\citep{Kaifu2004}.

\begin{figure}[]
\centering
\includegraphics[width=0.83\columnwidth,angle=0]{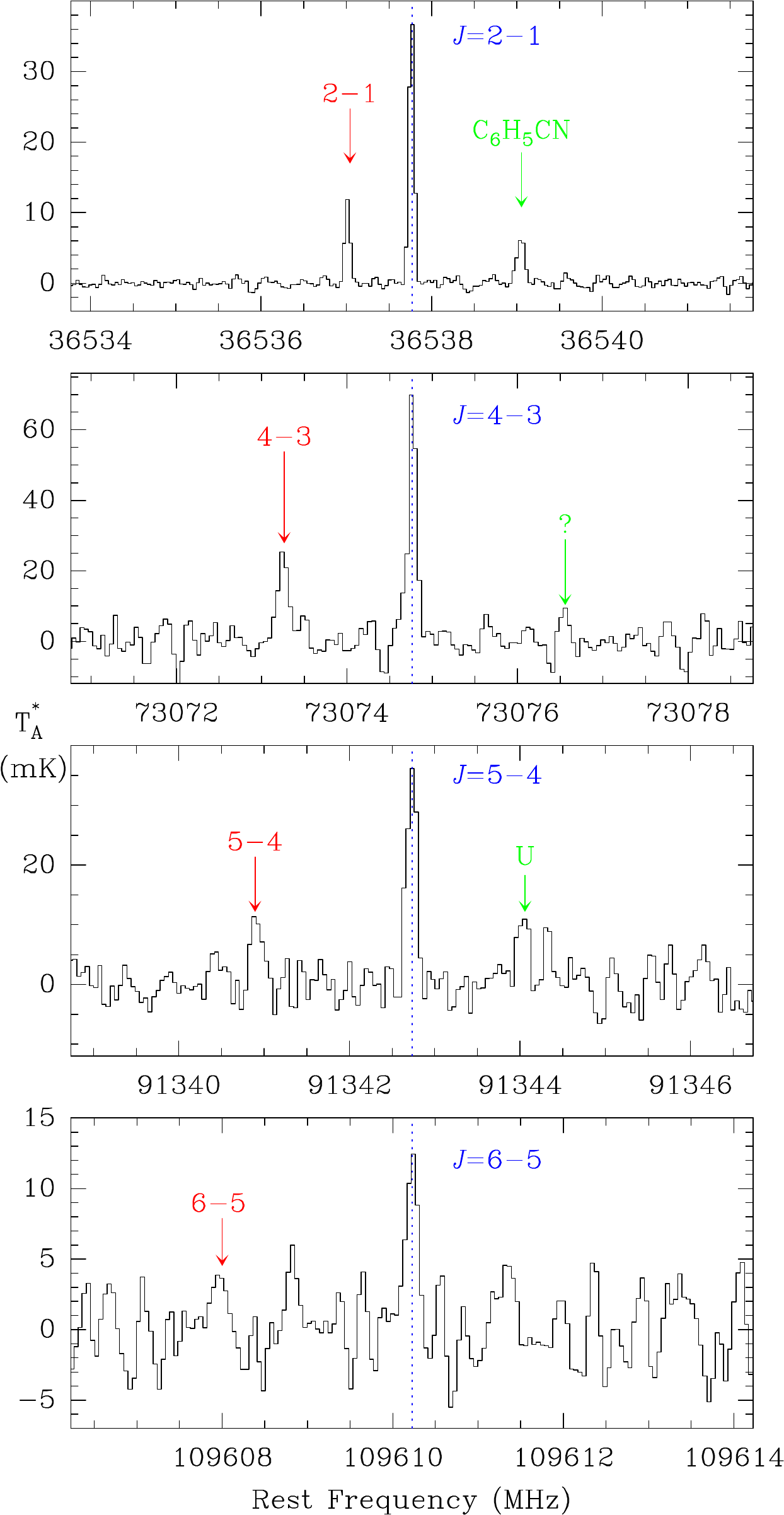}
\caption{Observed lines of CH$_3$CO$^+$ towards TMC-1.
The abscissa corresponds to rest frequencies (in MHz) assuming a local standard of rest velocity
of 5.83 km s$^{-1}$ \citep{Cernicharo2020a,Cernicharo2020b}. Frequencies and intensities for the
observed lines are given in Table \ref{tab_CH3CO+}.
The ordinate is the antenna temperature (in mK).
Spectral resolution is 38.1 kHz below 50 GHz and 48.8 kHz above. The blue labels correspond to the
series of lines we assign to the $A$ species of
CH$_3$CO$^+$, while the red ones correspond to those of the $E$ species.
}
\label{fig_ch3co+}
\end{figure}

The IRAM 30m data come from a line survey performed towards TMC-1 and B1, and the observations
have been described by \citet{Marcelino2007} and \cite{Cernicharo2012b}.
The observations of L1527 and L1544 were obtained as part of the IRAM 30m Large Program ASAI and were described by 
\citet{Lefloch2018}.
The intensity scale and antenna temperature
($T_A^*$) for the two telescopes used in this work were calibrated using two absorbers at different temperatures as well as the
atmospheric transmission model ATM \citep{Cernicharo1985, Pardo2001}.
Calibration uncertainties were adopted to be 10~\%.
All data were analysed using the GILDAS package\footnote{\texttt{http://www.iram.fr/IRAMFR/GILDAS}}.

\begin{table}
\tiny
\caption{Observed line parameters for CH$_3$CO$^+$ in TMC-1.}
\label{tab_CH3CO+}
\centering
\begin{tabular}{{ccccccc}}
\hline
{\textit J$_u$}& $K$& $\nu_{obs}^a$& $\Delta\nu_{oc}^b$  & T$_A^*$$^c$& $\Delta$v$^d$   &$\int$T$_A^*$dv $^e$\\
               &       &  (MHz)       &     (kHz)            & (mK)   & (km\,s$^{-1}$)  &(mK km\,s$^{-1}$)\\
\hline
2& 0 &    36537.765& -3.5& 39.0$\pm$0.6 & 0.63$\pm$0.01& 26.1$\pm$0.6\\
2& 1 &    36537.014& -0.6& 11.8$\pm$0.6 & 0.60$\pm$0.03&  7.5$\pm$0.6\\
4& 0 &    73074.769&  2.7& 71.0$\pm$3.5 & 0.47$\pm$0.03& 35.7$\pm$2.0\\
4& 1 &    73073.252& -6.6& 25.0$\pm$3.5 & 0.65$\pm$0.11& 17.2$\pm$2.0\\
5& 0 &    91342.732& -3.5& 37.5$\pm$3.0 & 0.46$\pm$0.04& 18.2$\pm$1.0\\
5& 1 &    91340.865& 14.2& 11.9$\pm$3.0 & 0.49$\pm$0.11&  6.3$\pm$1.0\\
6& 0 &   109610.225&  2.1& 12.8$\pm$2.8 & 0.45$\pm$0.09&  6.1$\pm$1.0\\
6& 1 &   109607.954& -7.3&  6.6$\pm$2.8 & 0.60$\pm$0.10&  4.2$\pm$1.0\\
\hline
\end{tabular}
\tablefoot{
        \tablefoottext{a}{Observed frequencies (in MHz) adopting a v$_{LSR}$ of 5.83 km s$^{-1}$ for TMC-1. The uncertainty is 10 kHz
    for all the lines.}
        \tablefoottext{b}{Observed minus calculated frequencies (in kHz) resulting from a fit to the
    observed frequencies. The $J$=7-6 $K$=0,1 lines observed in L1527 have been
    included in the fit (see text).}
        \tablefoottext{c}{Antenna temperature (in mK).}
        \tablefoottext{d}{Linewidth at half intensity 
   (in km\,s$^{-1}$).}
        \tablefoottext{e}{Integrated line intensity (in mK km\,s$^{-1}$).}
}
\end{table}
\normalsize

\section{Results and discussion}
\label{sec:results}
The assignment of the observed features in our line surveys was done using the CDMS and JPL catalogues \citep{Muller2005,Pickett1998}
and the MADEX code \citep{Cernicharo2012}.
Most of the weak lines found in our survey of TMC-1 can be assigned to known species and their isotopologues.
Nevertheless, many features remain unidentified. Frequencies for the unknown lines were derived by assuming a local standard of rest
velocity of 5.83 km s$^{-1}$, a value that was derived from the observed transitions
of HC$_5$N and its isotopologues in
our line survey \citep{Cernicharo2020a,Cernicharo2020b}.
Our new data towards TMC-1 allowed us to detect
C$_3$N$^-$ and C$_5$N$^-$ \citep{Cernicharo2020a}, as well as new species such as
the isocyano isomer of HC$_5$N, HC$_4$NC \citep{Cernicharo2020b},
the cation HC$_3$O$^+$ \citep{Cernicharo2020c},
the cation HC$_3$S$^+$ \citep{Cernicharo2020d}, and
the cation HC$_5$NH$^+$ \citep{Marcelino2020a},
in addition to several tens of already known molecules and their isotopologues.

Within the unidentified features in our surveys in the 3\,mm band and the Q-band, we found two series of four lines
with a harmonic relation of 2:4:5:6 (see Fig. \ref{fig_ch3co+}). Taking into account the line density in TMC-1,
the possibility that the observed pattern is fortuitous is very small.
The observed lines are shown in
Fig. \ref{fig_ch3co+}, and the derived line parameters are given in Table \ref{tab_CH3CO+}.
In fact, the $J$=5-4 line at 91342 MHz has intrigued us since 2017 when we detected it
in TMC-1, L483, L1527, and L1544. We interpreted the $K$=0,1 lines and the U line
at $\sim$91344 (see Fig. \ref{fig_other54}) as the hyperfine structure of a $J$=1-0
or $J$=2-1 transition of a molecule
containing a nucleus with a spin of 1. Using the old receivers of the Yebes 40 m telescope,
and assuming that the three lines around 91342 MHz could correspond to $J$=2-1, we searched for lines at
45671 MHz without success. Only when the new receivers covering the whole Q-band were available at the telescope, 
and we detected the doublet at 36537 MHz (see Fig. \ref{fig_ch3co+}), did we realized that two of the lines around 
91342 MHz correspond to a $J$=5-4 transition
in harmonic relation 2:5 with the 36537 MHz doublet. Moreover, the U line at 91344 MHz is produced by another carrier 
as it is detected in B1, while the other lines are not. Once we relaxed the
initial idea that these features were the hyperfine structure of
a low-$J$ transition, other features were found in the 3\,mm domain ($J$=4-3 and $J$=6-5, as well as $J$=7-6 in L1527).

The two series of lines can be fitted to two linear rotors with rotational constants
$B$ = 9134.4738 $\pm$ 0.0006 MHz and $B$ = 9134.2860 $\pm$ 0.0020 MHz. The distortion constant is exactly the
same for both series, $D$ = 4.00 $\pm$ 0.02 kHz. The observed spectra is reminiscent of the $K$=0 and $K$=1
components of the rotational transitions of a
symmetric rotor. In fact, the eight observed lines in TMC-1 can be fitted with a single rotational
constant and two distortion constants if we assume that the carrier is the same for both
series and that it has a $C_{3v}$ symmetry (i.e. that it is a symmetric rotor).
Using
the standard Hamiltonian for this kind of molecular rotor \citep{Gordy1984},
we derived the rotational and distortion constants provided in Table \ref{tab_fits}.

\begin{table}
\tiny
\caption{Derived spectroscopic parameters (in MHz) for CH$_3$CO$^+$.}
\label{tab_fits}
\centering
\begin{tabular}{{lccc}}
\hline
Constant          &  Space$^a$              & Laboratory$^b$             & Merged$^c$ \\
\hline
$B$               &9134.47424(82)           & 9134.47083(27)           & 9134.47211(20) \\
$D_J$             &   4.014(12) 10$^{-3}$  &    3.99198(25) 10$^{-3}$&    3.99307(21) 10$^{-3}$\\
$D_{JK}$          &   1.8847(53)10$^{-1}$  &    1.87616(41) 10$^{-1}$&    1.87736(46) 10$^{-1}$\\
$H_{JK}$          &                         &    8.66(33) 10$^{-7}$   &    9.56(37) 10$^{-7}$\\
$H_{KJ}$          &                         &    6.58(59) 10$^{-6}$   &    7.19(74) 10$^{-6}$\\
\hline
$rms$(kHz)$^d$    &   6.9                   &   34.3                     & 33.2  \\
$J_{min}/J_{max}$ &   1/7                   &   10/27                    &  1/27 \\
$K_{min}/K_{max}$ &   0/1                   &    0/6                     &  0/6  \\
$N_{lines}^e$     &   10                    &    79                      & 89\\
$\nu_ {max}$(GHz) & 127.87                  &   492.95                   & 492.95\\
\hline
\end{tabular}
\tablefoot{
        \tablefoottext{a}{Fit to the lines of CH$_3$CO$^+$ observed in TMC-1.
    In this fit, the $J$=7-6 and $K$=0,1 lines observed in L1527 (with frequencies of
    127877.133$\pm$0.025 and 127874.494$\pm$0.050 MHz, respectively) have been included
    (see Fig. \ref{fig_other54} and Appendix \ref{other_sources}).}
        \tablefoottext{b}{Fit to the lines of CH$_3$CO$^+$ observed in the laboratory.}
        \tablefoottext{c}{Fit to the the lines of CH$_3$CO$^+$ observed in space and in
    the laboratory.}
        \tablefoottext{d}{The standard deviation of the fit (in kHz).}
        \tablefoottext{e}{Number of lines included in the fit.}
    }
\end{table}
\normalsize

From the derived rotational constant, 9134 MHz, the molecule should contain at least
three atoms between C, N, and O. We analysed the possible candidates that could
have a rotational constant similar to the observed one. Detailed \textit{ab initio} calculations
for the possible linear and asymmetric carriers are given in Appendix \ref{potential}.
Concerning symmetric rotors, it is amazing to realize that all CH$_3$X, with X=CN, NC, and CCH,
have rotational constants close to our rotational and distortion constants. For example,
CH$_3$CN has a
rotational constant of 9198.9 MHz
\citep{Muller2009}, which is really very close to our result.
The other possible candidates, CH$_3$CNH$^+$ ($B$=8590.5 MHz;
\citealt{Amano2006}) and CH$_3$NCH$^+$ (see, Table \ref{abini_supp}), are too heavy.
Hence, the best symmetric rotor candidate seems to be a species similar to CH$_3$CN.
The acetyl radical, CH$_3$CO, has been observed in the laboratory by \citet{Hirota2006},
but it is asymmetric and its lines show a very complex hyperfine
structure. However, CH$_3$CO$^+$ is a symmetric rotor \citep{Mosley2014} and the lowest
energy isomer of H$_3$C$_2$O$^+$. Its possible precursor, if formed through protonation, is ketene,
which is one of the most abundant O-bearing species in TMC-1 (see \citealt{Cernicharo2020c}).

\begin{table}
\begin{tiny}
\caption{Scaled theoretical values for the spectroscopic parameters of CH$_3$CO$^+$, CH$_2$COH$^+$, and CH$_3$NCH$^+$ together with the experimental values obtained in this work (all in MHz).}
\label{exp_abini}
\centering
\begin{tabular}{{lccccc}}
\hline
\hline
Parameter & Exp.\tablefootmark{a} & CH$_3$CO$^+$ & CH$_2$COH$^+$ & CH$_3$NCH$^+$ \\
\hline
$B$                     &     9134.4742(8)    & 9129.6    &    9309.5\tablefootmark{b}    &     9105.5      \\
$D_J\times$10$^{-3}$    &     4.014(13)       & 3.9       &     2.8                       &     4.0         \\
$D_{JK}\times$10$^{-3}$ &    188.47(50)       & 184.6     &     378.7                     &     171.7       \\
\hline
\end{tabular}
\tablefoot{
\tablefoottext{a}{This work.}
\tablefoottext{b}{($B$+$C$)/2}
}
\end{tiny}
\end{table}

\subsection{Quantum chemical calculations and assignment to CH$_3$CO$^+$}
\label{ab_initio}
Precise geometries and spectroscopic molecular parameters for the species mentioned above were
computed using high-level \emph{ab initio} calculations. The first screening for all plausible candidates 
(see Appendix \ref{potential}) was done at the CCSD/cc-pVTZ level of theory
\citep{Cizek1969,Dunning1989}. These results are shown in Table \ref{abini_supp}. In a second stage, the 
most promising candidates, namely CH$_3$CO$^+$, CH$_2$COH$^+$, and CH$_3$NCH$^+$, were calculated at  the
CCSD(T)-F12b/aug-cc-pVQZ levels
of theory \citep{Raghavachari1989,Adler2007,Knizia2009}. To obtain more precise values for the rotational
parameters of these three species, we calibrated our calculations using
experimental to theoretical scaling ratios for analogue molecular species. This method has been proved to be suitable 
to accurately reproduce the molecular geometry of other identified molecules
\citep{Cernicharo2019,Marcelino2020a,Cernicharo2020c}. In our present case, we used CH$_3$CN, CH$_2$CNH, and CH$_3$NC, which are isoelectronic species of CH$_3$CO$^+$, CH$_2$COH$^+$, and CH$_3$NCH$^+$, respectively, for this
purpose. Table \ref{abini} shows the results of these calculations, which are summarized in Table \ref{exp_abini}. As can be seen, the employed level of theory reproduces the rotational parameters
for CH$_3$CN,  CH$_2$CNH, and CH$_3$NC very well, with relative discrepancies around 0.08\% and 0.04\% for $B$ in the cases of
CH$_3$CN and CH$_3$NC, respectively. After correcting the calculated parameters for CH$_3$CO$^+$, CH$_2$COH$^+$, and CH$_3$NCH$^+$ using 
the derived scaling ratios for CH$_3$CN, CH$_2$CNH, and CH$_3$NC, respectively, we obtained a $B$ constant of
9129.62 MHz for CH$_3$CO$^+,$ which shows the best agreement with that derived from the TMC-1 lines.
The centrifugal distortion values, obtained in the same manner but at the MP2/aug-cc-pVQZ level
of theory for CH$_3$CO$^+$ and CH$_3$NCH$^+$, are both compatible with those obtained
from the fit of the lines. The agreement between the experimental constants and those
calculated for CH$_2$COH$^+$ is substantially worse. The calculated dipole moments
for CH$_3$CO$^+$ and CH$_3$NCH$^+$ are 3.5 D and  2.0 D, respectively, while the
$\mu_a$ and $\mu_b$ values for CH$_2$COH$^+$ are 0.8 and 1.7 D, respectively.

In addition to the geometry optimizations, we calculated the energy associated with the plausible
formation of CH$_3$CO$^+$, starting from ketene and three proton donors; H$_3^+$, H$_3$O$^+$, and HCO$^+$. All these calculations were carried out at the CCSD/cc-pVTZ level of theory. 
We found a total energy change in the protonation of ketene to form CH$_3$CO$^+$
of $-$421.8, $-$130.9, and $-$244.8 kJ mol$^{-1}$ when ketene reacts with H$_3^+$, H$_3$O$^+$, and HCO$^+$,
respectively. More details can be found in Appendix \ref{more_calcu}.

\begin{figure}[]
\centering
\includegraphics[width=0.99\columnwidth,angle=0]{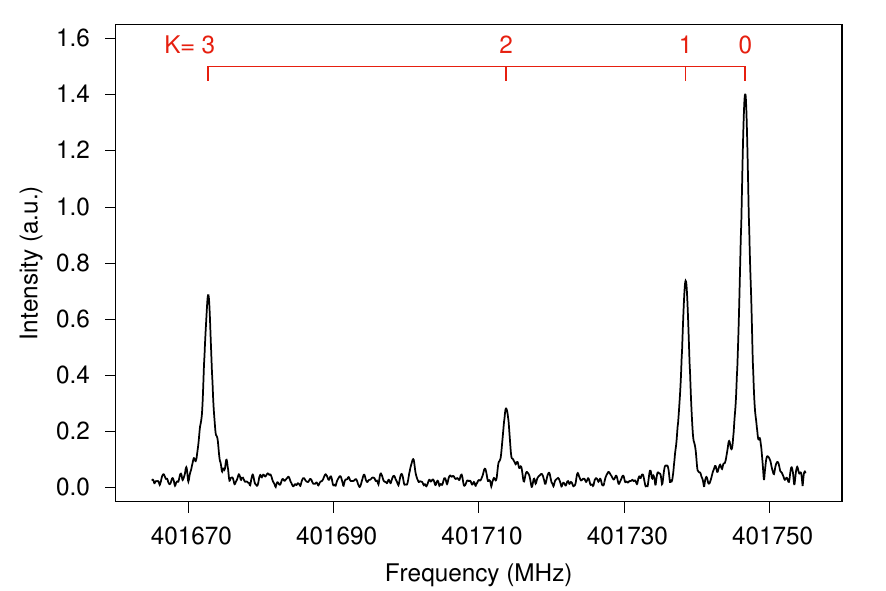}
\caption{ $J$=22$\rightarrow$21, $K$=0–3 transitions obtained by chirped-pulsed
excitation. The record corresponds to the average of 67 million spectra acquired in $\sim$20 min.}
\label{fig_labo}
\end{figure}

\subsection{Laboratory detection of CH$_3$CO$^+$}
\label{laboratory}

We conducted an experiment to detect the CH$_3$CO$^+$ cation in the laboratory using rotational spectroscopy below 500 GHz.
The experimental setup was similar to the one used to detect
NS$^+$ \citep{Cernicharo2018}. The cation was produced in a liquid-nitrogen-cooled Pyrex absorption cell by glow-discharging a mixture of
CH$_4$, CO (1:1), and Ar.
A solenoid coil wound on the cell can generate an axial magnetic
field (up to 300 G) to magnetically extend the negative glow, the region known
to produce the highest concentrations of cations (compared to the positive
column discharge; \citealt{DeLucia1983}).
We also tried acetone and
acetaldehyde as precursors \citep{Mosley2014}, but without success.

To optimize the experimental setup, we first observed the $J$=2$\leftarrow$1 transition
of HCO$^+$ at 178375.056 MHz, which was produced in the same gas mixture.
We then searched for the $J$=10$\leftarrow$9, $K$=0-2 series of lines of CH$_3$CO$^+$ between 182.658 and 182.675 GHz
based on the rotational constants derived from the lines observed in TMC-1.
Weak spectra were observed within 500 kHz.
The best experimental conditions were found to be P(CH$_4$)=P(CO)=1.5 mTorr, P(Ar)=5.5 mTorr
(gas mixture cooled using liquid nitrogen but pressures measured at room temperature),
an electric discharge of 3.5 kV/10 mA, and an axial magnetic field of 200 G.
These lines disappeared when one of the
precursors was suppressed, or when the axial magnetic
field was cut off. The latter phenomenon confirmed almost unambiguously that they were due
to a cation. Subsequent measurements of higher-$J$ transitions fully
support the astrophysical assignment of the observed lines to CH$_3$CO$^+$.

In total, 79 lines were observed in the laboratory with quantum numbers in the ranges
$J$=10-27 and $K\leq$6 (see Table \ref{freq_labo}). Transitions occurring below 330 GHz were measured by standard frequency
modulation absorption spectroscopy, resulting in second-derivative lineshapes.
These lines ($K\leq$3) were found unshifted from the first prediction.
Those from 400 to 500 GHz were measured by emission spectroscopy \citep{Zou2020}, giving Voigt-profile
lineshapes. Compared to the prediction, some deviations were observed up
to 1 MHz for $K$=6; these measurements led us to determine the $H_{JK}$ and $H_{KJ}$
centrifugal distortion terms. For maximum sensitivity, these lines were measured
using the single frequency excitation method with 5-20 million acquisitions
(which took 1 to 5 min.). Additionally, a 120 MHz wide chirped excitation spectrum,
measured with 67 million acquisitions, is given in Fig. \ref{fig_labo}
for illustration and comparison purposes.
The uncertainty of the laboratory frequency measurements are estimated to be 50 kHz.
Given the mass of the cation, and that the negative glow is a nearly electric
field-free region, the reported laboratory frequencies are expected to be
unshifted by the Doppler effect.
The separate and merged least-squares analysis of all (astronomical and laboratory) measured
transitions are provided in Table \ref{tab_fits}. The measured frequencies and the observed minus calculated
values are given in Table \ref{freq_labo}. Frequency predictions are given in Table \ref{freq_pred}.

\subsection{Chemistry of CH$_3$CO$^+$}
\label{dis_chem}
From the observed line intensities of CH$_3$CO$^+$, we derived a rotational temperature of $\sim$5 K and a total column density of 
(3.2$\pm$0.3)$\times$10$^{11}$ cm$^{-2}$ (see Appendix \ref{CH3_X}). 
The column densities for the $A$ and $E$ species are (2.2$\pm$0.2)$\times$10$^{11}$ cm$^{-2}$
and (9.7$\pm$0.9)$\times$10$^{10}$ cm$^{-2}$, respectively. 
Adopting the column density for
ketene derived by \citet{Cernicharo2020c}, we obtained a H$_2$CCO/CH$_3$CO$^+$ ratio of 44. Assuming
the H$_2$ column density derived by \citet{Cernicharo1987}, the abundance of CH$_3$CO$^+$ is
3.2$\times$10$^{-11}$.

The chemistry of protonated molecules in cold dense clouds has been discussed by
\cite{Agundez2015}. Chemical model calculations similar to those that they presented predict that the abundance of protonated ketene is controlled by the typical
routes operating for protonated molecules. That is, CH$_3$CO$^+$ is mostly formed
by proton transfer to H$_2$CCO from HCO$^+$, H$_3$$^+$, and H$_3$O$^+$, while it is
destroyed through dissociative recombination with electrons. The radiative association
between CH$_3^+$ and CO is also an important route to CH$_3$CO$^+$. The abundance ratio
H$_2$CCO/CH$_3$CO$^+$ predicted by the model is in the range 250-450 and depends on whether
the UMIST {\small RATE12} \citep{McElroy2013} or
KIDA {\small \texttt{kida.uva.2014}} \citep{Wakelam2015} chemical networks are used.
As occurs for most protonated molecules observed in cold dense clouds, the abundance of
the protonated form with respect to the neutral is underestimated by the chemical model.
In this case, there is a factor of 5-10 difference between the model and observations.
Incorrect estimates for the rate constants of the dominant reactions of the formation and
destruction of CH$_3$CO$^+$ may be behind this disagreement. Alternatively, the chemical
network may miss some important formation route to CH$_3$CO$^+$, although it is difficult
to identify reactions producing this ion from abundant reagents. For example, plausible
reactions of CH$_n^+$ ions with CO, H$_2$CO, or CH$_3$OH tend to form products other than
CH$_3$CO$^+$ \citep{Adams1978}.
In this context, it is worth noting that not all species resulting from the protonation
of abundant molecules in TMC-1 are detected. For example, CH$_3$CNH$^+$ is
not detected in TMC-1 despite the CH$_3$CN proton affinity of 787.4$\pm$5.9
kJ mol$^{-1}$  \citep{Williams2001}. The 3$\sigma$ upper limit to the column density of CH$_3$CNH$^+$
is $2.5\times10^{11}$ cm$^{-2}$. The column density of CH$_3$CN is $(3.2\pm0.2)\times10^{12}$ cm$^{-2}$ (see
Appendix \ref{other_sources}); hence, the abundance ratio between the neutral and its protonated form is $\ge$13.
The low dipole moment of CH$_3$CNH$^+$ compared to that of CH$_3$CN (1.01\,D versus 3.93\,D) limits
the chances of detecting this species.

\subsection{$A$-CH$_3$CO$^+$/$E$-CH$_3$CO$^+$ abundance ratio}
\label{interconversion}
The column densities derived for the $A$ and $E$ species of CH$_3$CO$^+$ are not identical,
as would be expected for a symmetric top. The $A$/$E$ abundance ratio for this molecule is 2.27.
However, all symmetric molecules of CH$_3$X detected in TMC-1
have an abundance ratio between their $A$ and $E$ species that is close to unity (see Appendix \ref{CH3_X} and
Fig. \ref{ae_rotor}). In a symmetric top, the two symmetry states $A$ and $E$  are not connected radiatively 
nor through inelastic collisions 
with H$_2$. Unlike the rest of the CH$_3$X molecules detected in TMC-1, CH$_3$CO$^+$ is a cation, and its reactive collisions
with H$_2$ or H could produce a proton interchange if there is no barrier to the reaction.
The lowest energy
level of the $E$ symmetry state is the $J$=1, $K$=1, which is 7.8 K above the ground $J$=0, $K$=0 level of
the $A$ state. Hence, the reaction of interchange of a proton
\\
\\
\noindent
$E$-CH$_3$CO$^+$ + H$_2$/H $\rightarrow$ $A$-CH$_3$CO$^+$ + H$_2$/H + 7.8 K
\\
\\
\noindent
is exothermic, although it is unknown if there is a barrier; this is something that has to be established via detailed theoretical calculations. At thermal equilibrium, and for a kinetic temperature of 10 K, the $A$/$E$ abundance
ratio could be $e^{0.78}$=2.18, which is very close to the observed value of 2.27. For neutral
molecules with two or more symmetric hydrogens, the proton interchange could be mainly produced through
collisions with H$^+$, H$_3$$^+$, HCO$^+$, and H$_3$O$^+$, which
are much less abundant than H$_2$ and H. In Appendix \ref{CH3_X}, we discuss
the $A/E$ abundance ratio of all neutral symmetric rotors that have been detected so far in TMC-1,
including CH$_3$NC, which has previously only been  observed in two cold dense clouds: L1544 \citep{Jimenez-Serra2016} and L483 \citep{Agundez2019}. For all these species,
the $A$/$E$ abundance ratio is close to unity.

Alternatively, we could also consider the
possibility that the collisional rates of the acetyl cation with H$_2$ or He are higher for the
$A$ species than for the $E$ species. As the acetyl cation is isoelectronic to CH$_3$CN, we could
use the collisional rates of the latter species \citep{Khalifa2020} to estimate possible differences
in the excitation temperature of the $K$=0 and $K$=1 lines. We explored a density range
of $(4-10)\times10^4$ cm$^{-3}$ and a kinetic temperature range of 5-10 K. No significant differences
were found in the predicted brightness temperature between these lines. Of course, if the effect is due to inelastic collisions, then methyl cyanide (CH$_3$CN) would
also show a similar behaviour. Nevertheless, although both species are isoelectronic, the fact that
CH$_3$CO$^+$ is positively charged could result in very different collisional rates with H$_2$ compared
to CH$_3$CN.

We could also consider that the $A/E$ abundance ratio
is affected by the formation process of the molecule. As shown in Sect. \ref{ab_initio}, the
reaction of ketene with H$_3$$^+$ is the most favourable for protonation from the thermodynamical
point of view. Both species, ketene and H$_3$$^+$, could also have their ortho/para ratio
affected by the low temperature of dense dark clouds, which will introduce a
non-trivial spin statistic into the formation process of CH$_3$CO$^+$.
Additional calculations are needed to evaluate the
role of collisional excitation and of spin interchange in order to understand the anomalous
behaviour exhibited by the $A$ and $E$ symmetry species of CH$_3$CO$^+$.

\begin{acknowledgements}
The Spanish authors thank Ministerio de Ciencia e Innovaci\'on for funding
support through project AYA2016-75066-C2-1-P, PID2019-106235GB-I00 and
PID2019-107115GB-C21 / AEI / 10.13039/501100011033. We also thank ERC for funding through grant
ERC-2013-Syg-610256-NANOCOSMOS. MA and CB thanks Ministerio de Ciencia e Innovaci\'on
for grants RyC-2014-16277 and FJCI-2016-27983, respectively. Y. Endo thanks Ministry of Science
and Technology of Taiwan through grant MOST108-2113-M-009-25. We would like to thank Evelyne Roueff 
and Octavio Roncero for useful comments and suggestions.
\end{acknowledgements}

\normalsize

\begin{appendix}
\section{CH$_3$CO$^+$ in other sources}
\label{other_sources}
\begin{figure}[]
\centering
\includegraphics[width=0.85\columnwidth,angle=0]{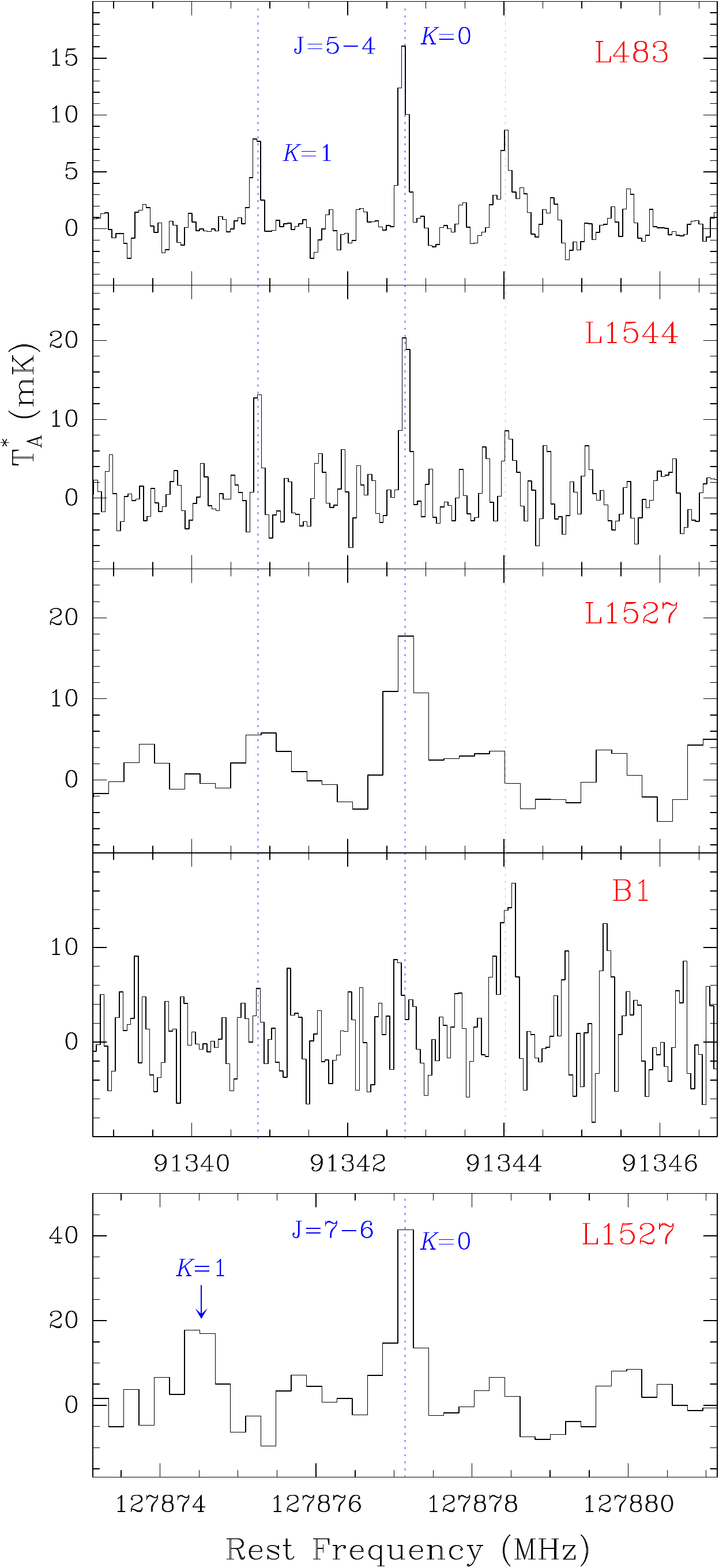}
\caption{Observations of the $J$=5-4 transition of CH$_3$CO$^+$ towards
L483, L1544, L1527, and B1 (top panels); the bottom panel
shows the $J$=7-6 line towards L1527.
The abscissa corresponds to the rest frequency (in MHz) and the ordinate is the antenna temperature (in mK).
The spectral resolution is 48.8 kHz for all sources except L1527, for which it is 198 kHz. The vertical dashed
blue lines indicate the position of the $K=0$ and $K=1$ lines (detected in all sources except B1),
and the cyan line corresponds to the U feature at
91344 MHz (detected in all sources except L1527).
The rest velocities of L1527 and L1544 were taken as 
v$_{LSR}$= 5.9 km/s and 7.2 km/s, respectively, based on \citet{Sakai2009} and \citet{Vastel2015}, respectively.
}
\label{fig_other54}
\end{figure}

The acetyl cation, CH$_3$CO$^+$,  has also been detected towards L483, L1544, and L1527 (see Fig. \ref{fig_other54}).
However, it is not observed towards Sgr B2 (PRIMOS
\footnote{
Access to the entire PRIMOS data set, specifics on the observing
strategy, and overall frequency coverage information is available at
http://www.cv.nrao.edu/~aremijan/PRIMOS/} line survey; \citealt{Neill2012}),
towards Orion-KL \citep{Tercero2010,Tercero2011},
or in our line survey of B1 \citep{Marcelino2007,Marcelino2009,Marcelino2010,
Cernicharo2012}. In the PRIMOS data on SgrB2, a very tentative detection
of the $J$=1-0 $K$=0 line could be claimed at a velocity of 80 km\,s$^{-1}$. However, only
an upper limit can be obtained for the $J$=2-1 transition as this line is heavily blended
with a strong line of acetone. It seems, hence, that CH$_3$CO$^+$ is typical of cold interstellar
clouds. 

\section{Potential carriers of the series of lines}
\label{potential}
All known diatomic species in cold dark clouds
have overly large rotational constants compared with those derived from the lines in TMC-1. For example,
a molecule containing one S
(CS, NS, SO) will have a rotational constant that is too high by more than 12 GHz.
Adding one or two H atoms to these
combinations of sulphur produces radicals (with overly high rotational constants, such as HCS) or asymmetric rotors such as H$_2$CS, which is too heavy \citep{Muller2019}.
A molecule with two S atoms is, of course, too heavy (for example,
$B$(S$_2$)=8831 MHz; \citealt{Pickett1979}), and we do not expect to have
Si- or P-bearing polyatomic species in this cloud. The first step in finding candidates is
to exclude the possibility of having a slightly asymmetric species. In that case, we could expect to have
lines corresponding to $K$=$\pm$1 at roughly $\pm$($B$-$C$) from the $K$=0 lines.
We searched in the Q-band survey ($J$=2-1) for such a pattern. No lines up to one-fourth of the intensity of the $K$=0 line are observed. Moreover, taking into account that there is no evidence for
a radical as a possible carrier, the species resulting from the addition of one hydrogen
to the closed-shell asymmetric species HNCO, HCOOH, and H$_2$CCC have
to be excluded. However, their protonated species are also, at least in principle, closed-shell
species. Hence, possible candidates are
H$_2$CCCH$^+$, CH$_3$CCH$_2$$^+$, HCNOH$^+$, HNCOH$^+$, HCOOH$_2$$^+$, and CH$_3$OO$^+$,
all of which, with the exception of the last one, are protonated forms
of known neutral and abundant species in TMC-1. Nevertheless, the resulting
molecular structures will be highly asymmetric for most of them,
or they are too light or too heavy, as are the cases for
H$_2$CCCH$^+$ and CH$_3$CCH$_2$$^+$, respectively.  HNCOH$^+$ is
a linear species characterized in the laboratory with a rotational constant of
9955 MHz \citep{Latanzi2012} and is not detected in our data.
Other exotic species, such as NH$_2$CHOH$^+$, CH$_2$ONH$_2$$^+$,
and CH$_3$NOH$^+$, which could result from the protonation of interesting molecules (NH$_2$CHO
for example), are discarded for their molecular asymmetry and because the neutral
species have not been observed in TMC-1.
Ab initio calculations have
been performed for the most promising candidates (see Table \ref{abini_supp}), and their isomers
and the results are discussed in Sect. \ref{ab_initio}.

\begin{table*}
        \caption{Rotational constants and electric dipole moments of potential candidates.}
        \label{abini_supp}
        \centering
        \begin{tabular}{{ccccc}}
                \hline\hline
                Species\tablefootmark{(a)} & {$\Delta E$}/kJ/mol \tablefootmark{(b)} & {$A$, $B$, $C$}/MHz & {$D_J$, $D_{JK}$}/kHz &  {$\mu_a$, $\mu_b$, $\mu_c$}/D    \\ \hline
        \multirow{2}{*}{\includegraphics[height=0.5cm]{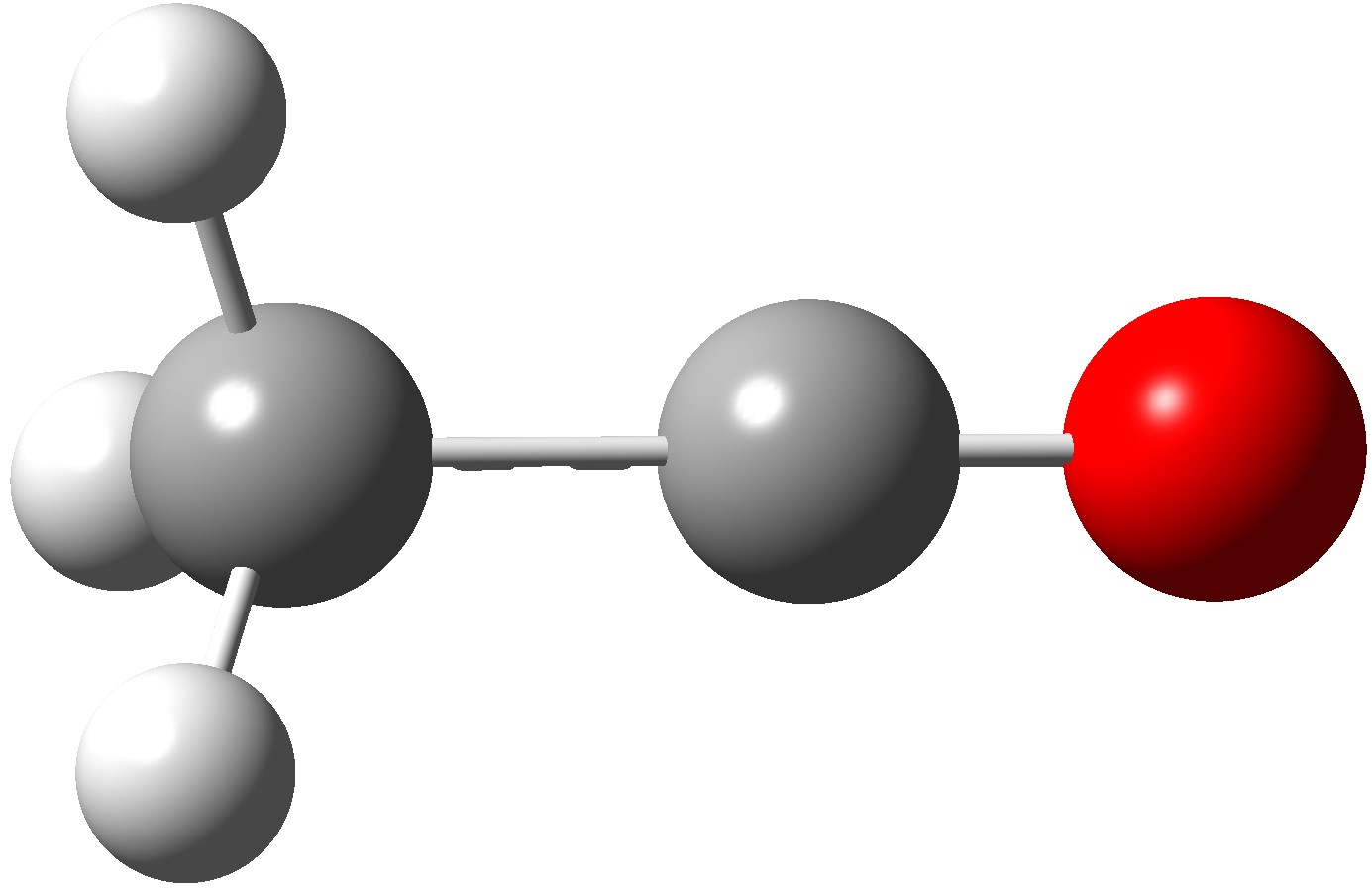}}   &        & 154355.3 & 3.9       & 2.5  \\
                                                                         & 0.0    & 9107.2   & 178.7     & 0.0  \\
    H$_{3}$CCO$^{+}$                                             &        & 9107.2   &           & 0.0  \\ \hline
    \multirow{2}{*}{\includegraphics[height=0.5cm]{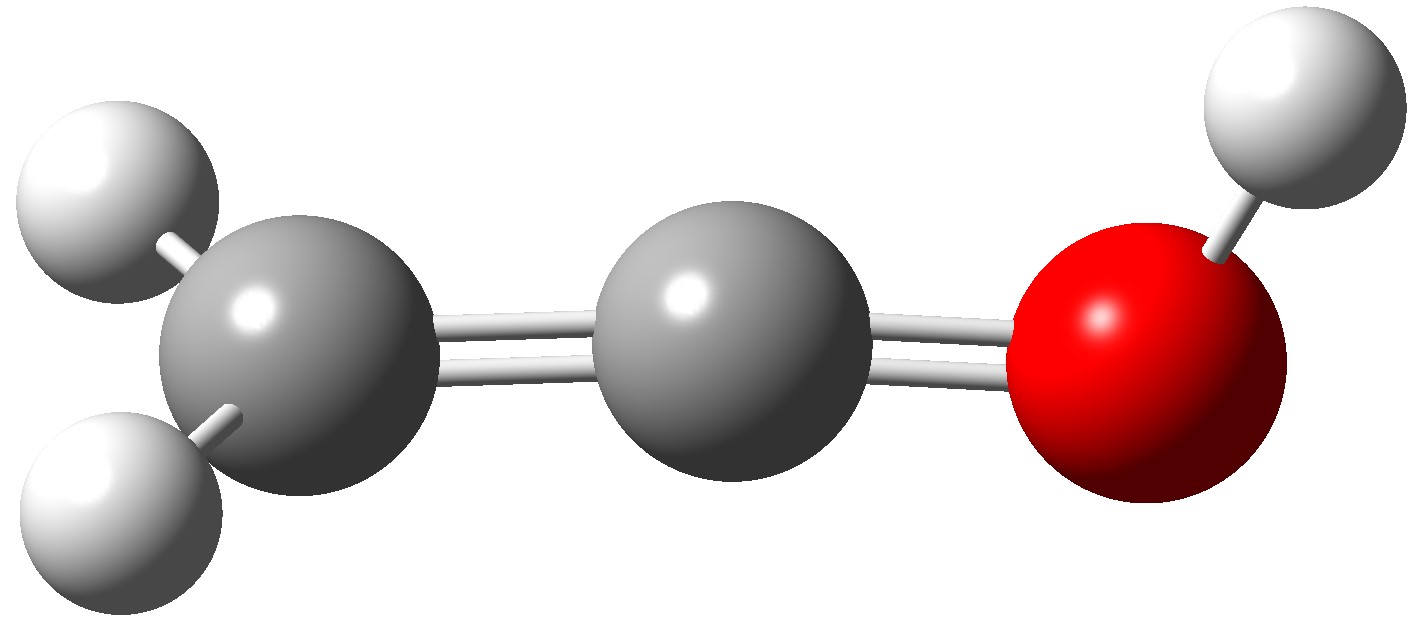}}  &        & 201325.3  & 2.7      & 1.6  \\
                                                                 & 178.0  & 9411.2   & 341.6     & 0.1  \\
        H$_{2}$COH$^{+}$                                             &        & 9222.3    &          & 0.0  \\ \hline \hline
    \multirow{2}{*}{\includegraphics[height=0.5cm]{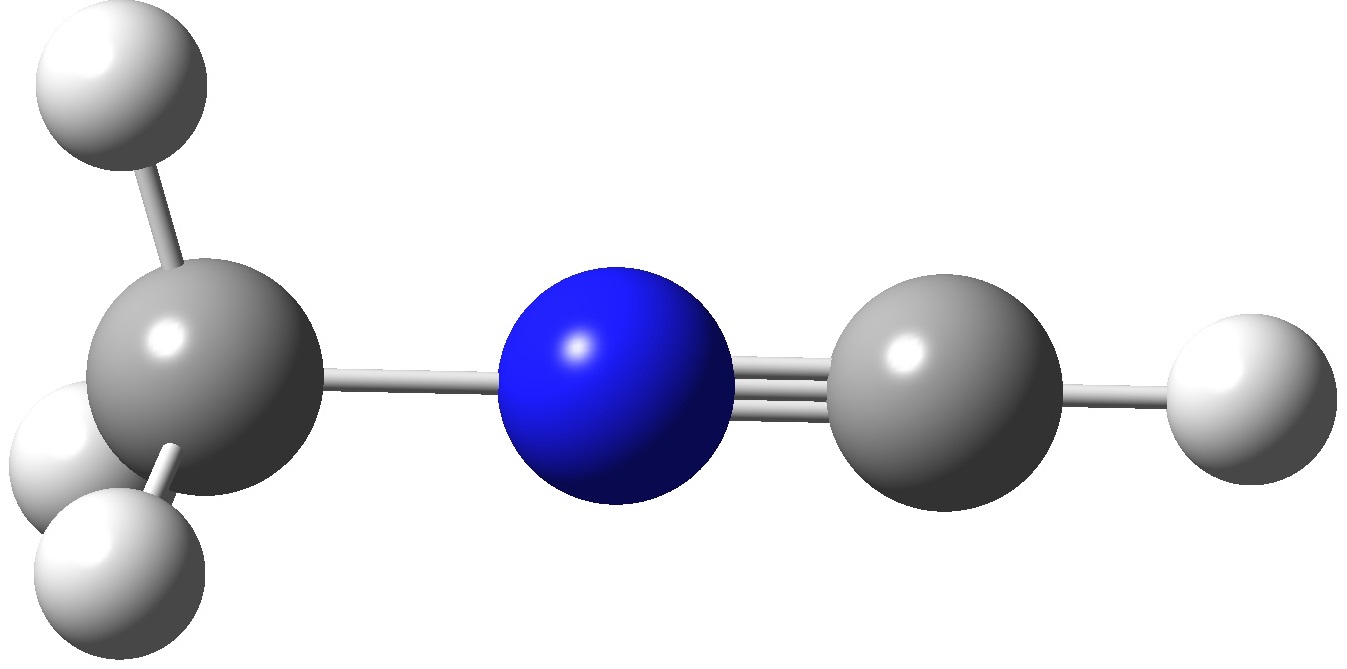}}  &        & 155571.2 & 3.8      & 1.9  \\
                                                                         &   0.0  & 9090.7   & 160.6     & 0.0  \\
        H$_{3}$CNCH$^{+}$                                            &        & 9090.7   &           & 0.0  \\ \hline \hline
    \multirow{2}{*}{\includegraphics[height=0.5cm]{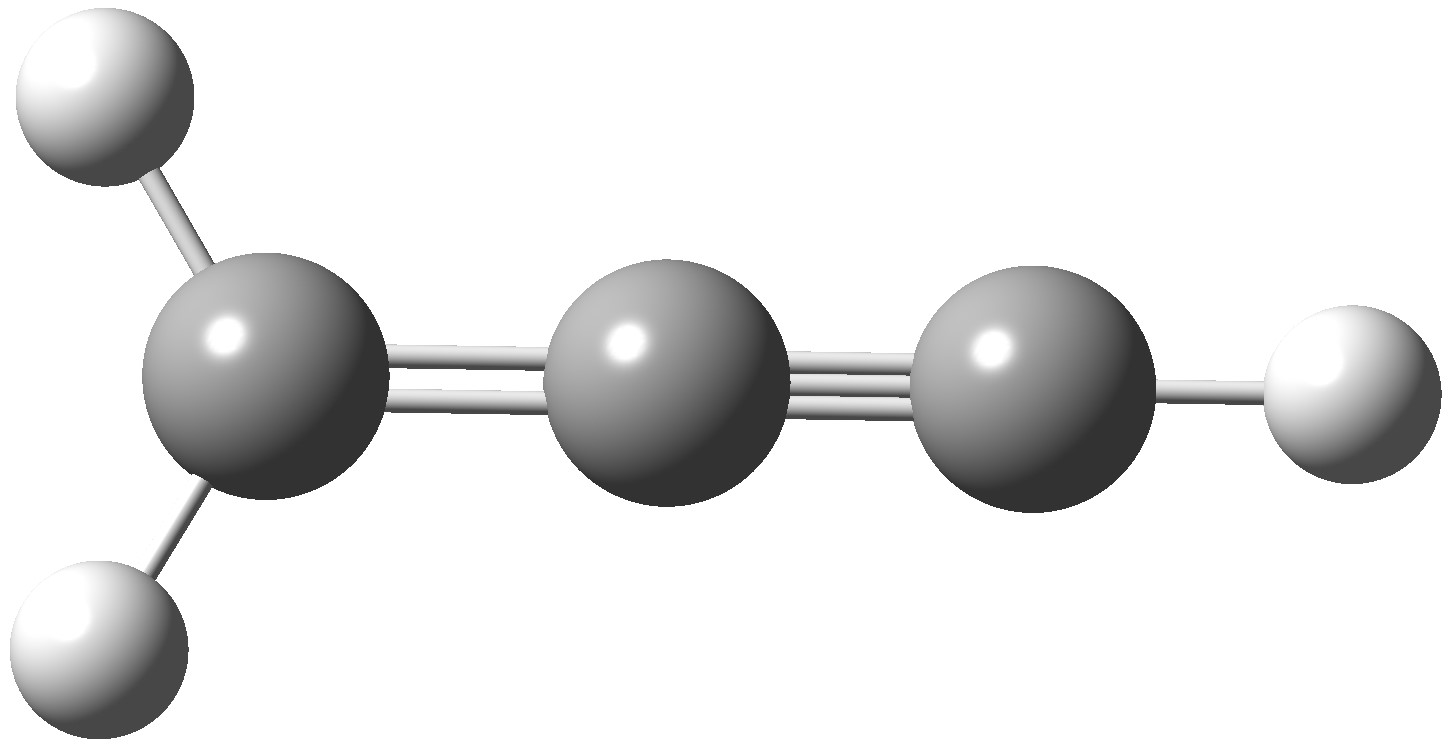}}  &        & 285625.5 & 2.6       & 0.5  \\
                                                                         & 0.0    & 9673.5   & 457.3     & 0.0  \\
        H$_{2}$CCCH$^{+}$                                            &        & 9356.6   &           & 0.0  \\ \hline \hline
        \multirow{2}{*}{\includegraphics[height=0.5cm]{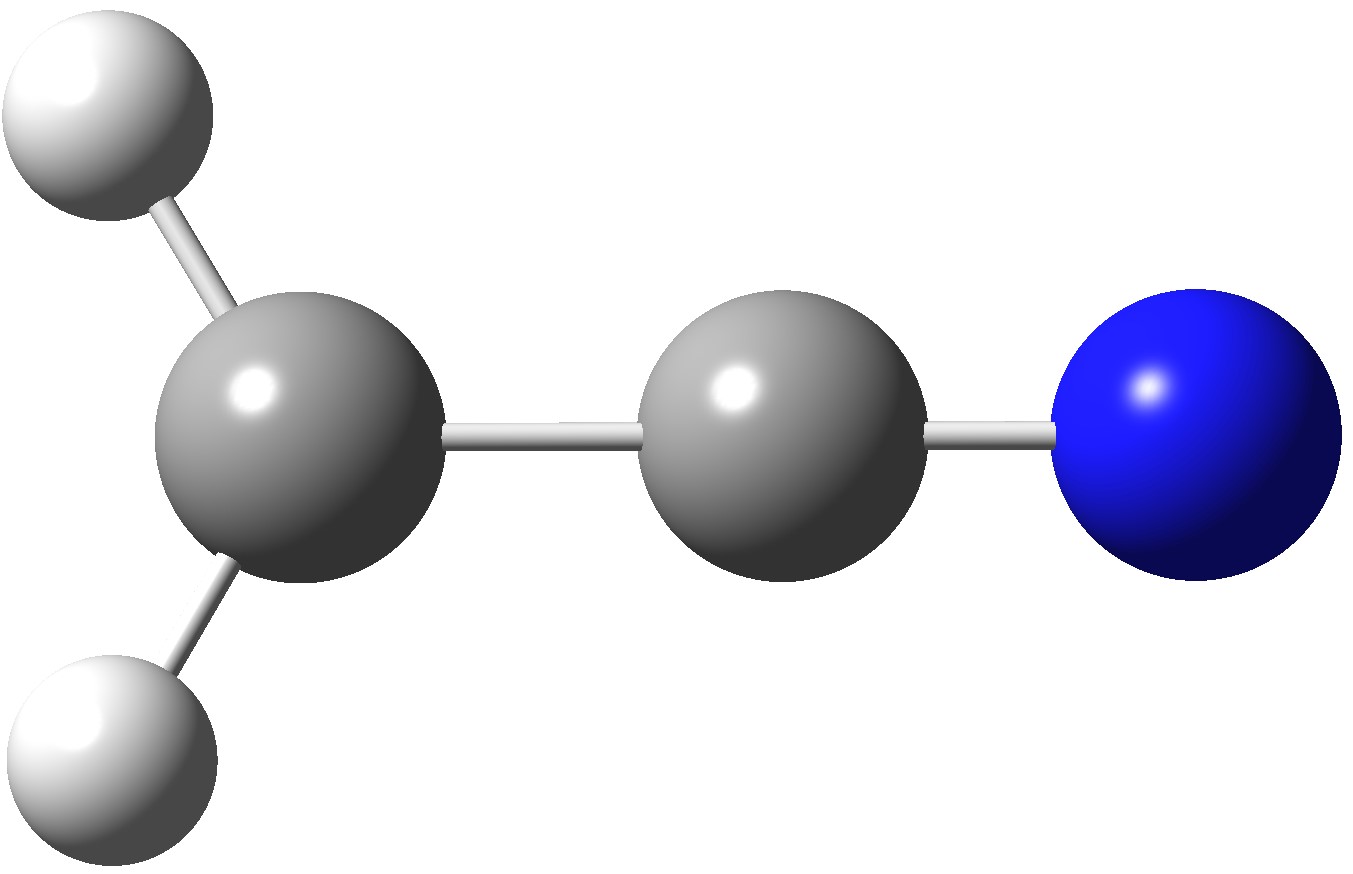}}   &        & 280905.8  & 3.5      & 5.0  \\
                                                                         & 0.0    & 10338.2   & 492.0    & 0.0  \\
        H$_{2}$CCN$^{+}$                                             &        & 9971.2     &         & 0.0  \\ \hline
        \multirow{2}{*}{\includegraphics[height=0.5cm]{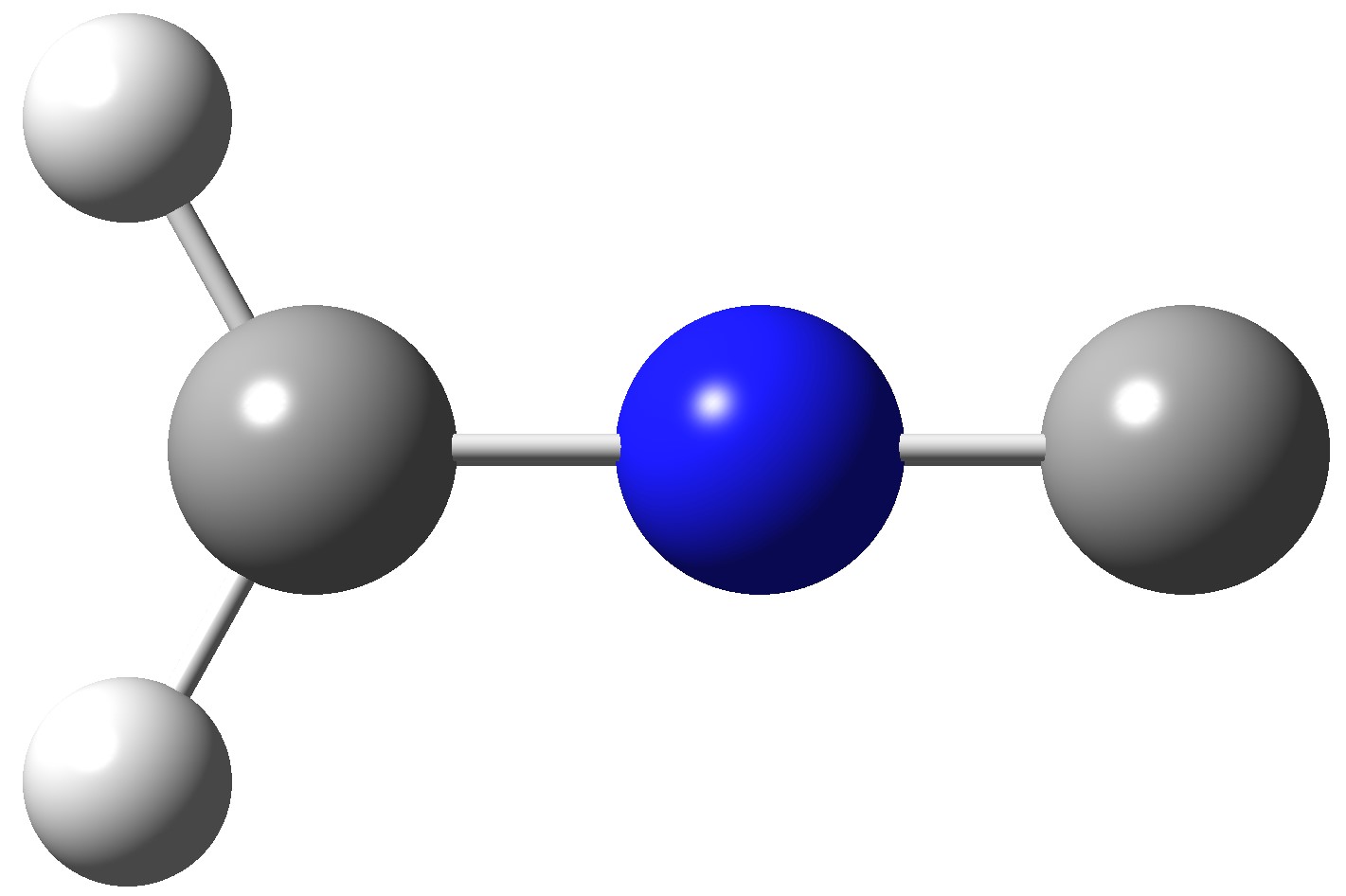}}   &        & 277331.4   & 4.00    & 3.7  \\
                                                                         &  5.9   & 11582.1    & 685.4   & 0.0  \\
        H$_{2}$CNC$^{+}$                                             &        & 11117.8    &         & 0.0  \\ \hline
        \multirow{2}{*}{\includegraphics[height=0.4cm]{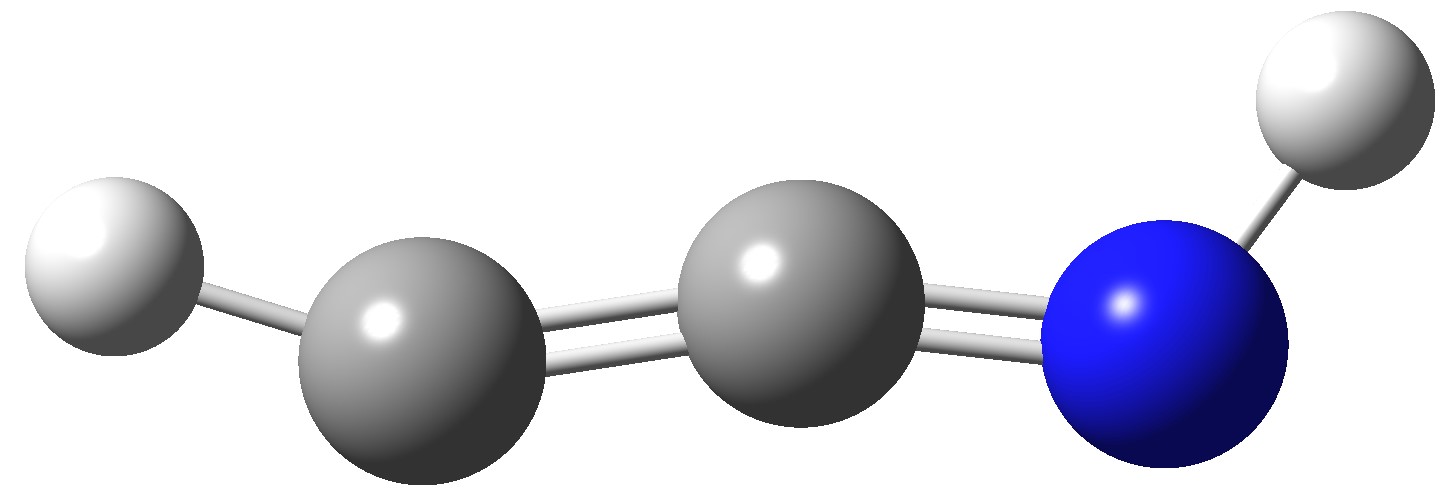}}   &        & 595150.6   & 2.7     & 1.5  \\
                                                                         &  102.5 & 10359.3    & 1285.8  &1.9   \\
        HNCCH$^{+}$                                                  &        & 10182.1    &         & 0.0  \\ \hline
        \multirow{2}{*}{\includegraphics[height=0.5cm]{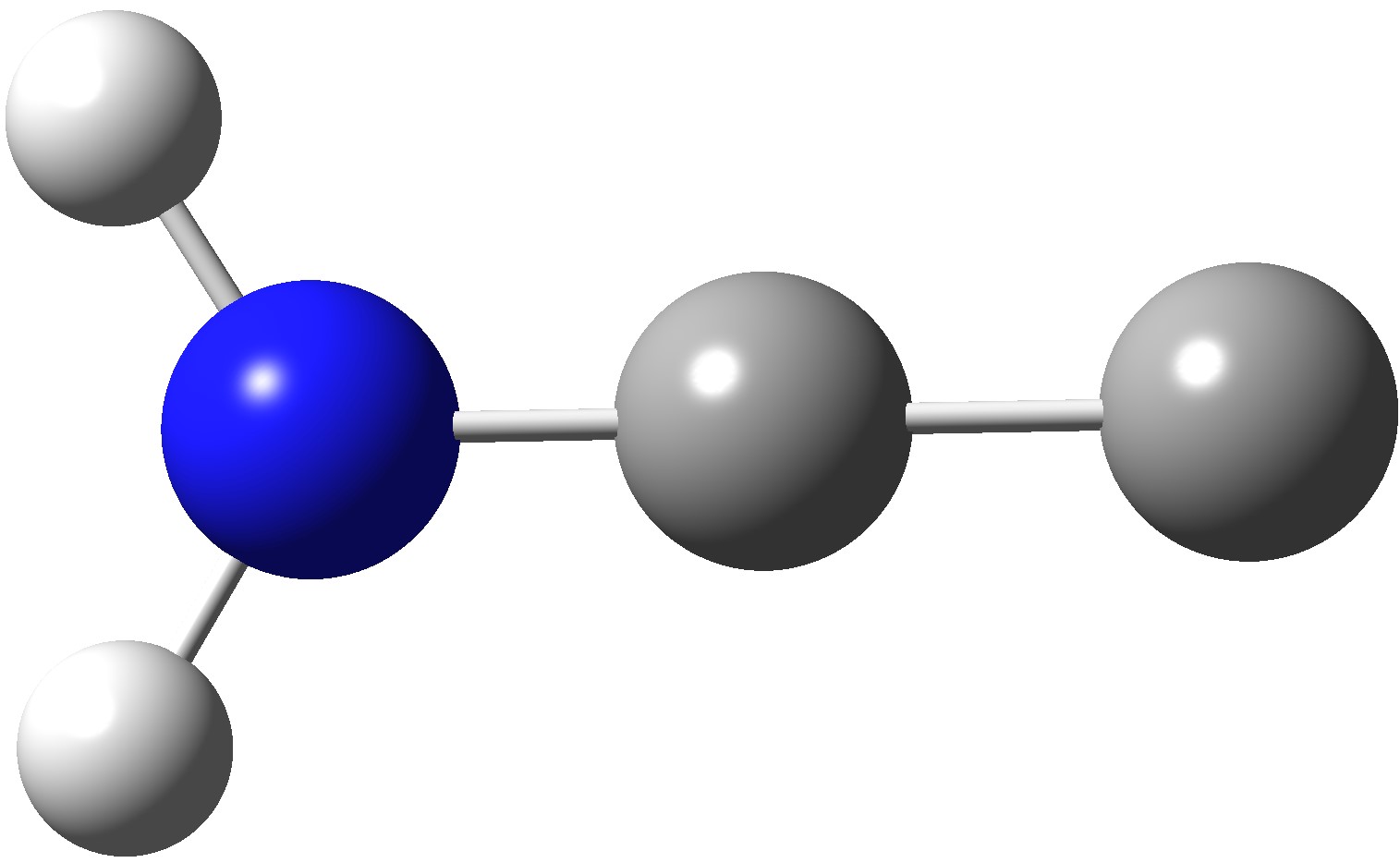}}   &        & 328396.0   & 3.6     & 4.6  \\
                                                                     &  176.4 & 10180.7    & 593.3   & 0.0  \\
        H$_{2}$NCC$^{+}$                                             &        & 9874.5     &         & 0.0  \\ \hline
        \multirow{2}{*}{\includegraphics[height=0.2cm]{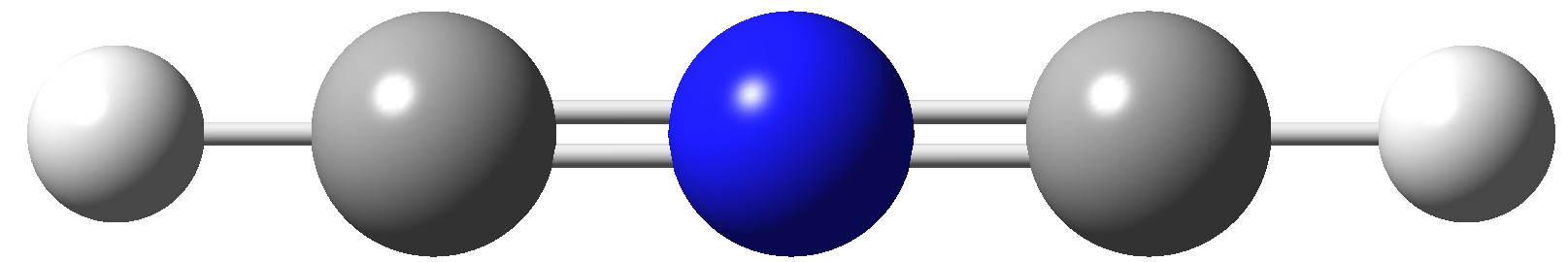}}   &        &            &         &      \\
                                                                     & 196.2  & 11120.9    & 2.9     & 0.0  \\
        HCNCH$^{+}$                                                  &        &            &         &      \\ \hline \hline
        \multirow{2}{*}{\includegraphics[height=0.5cm]{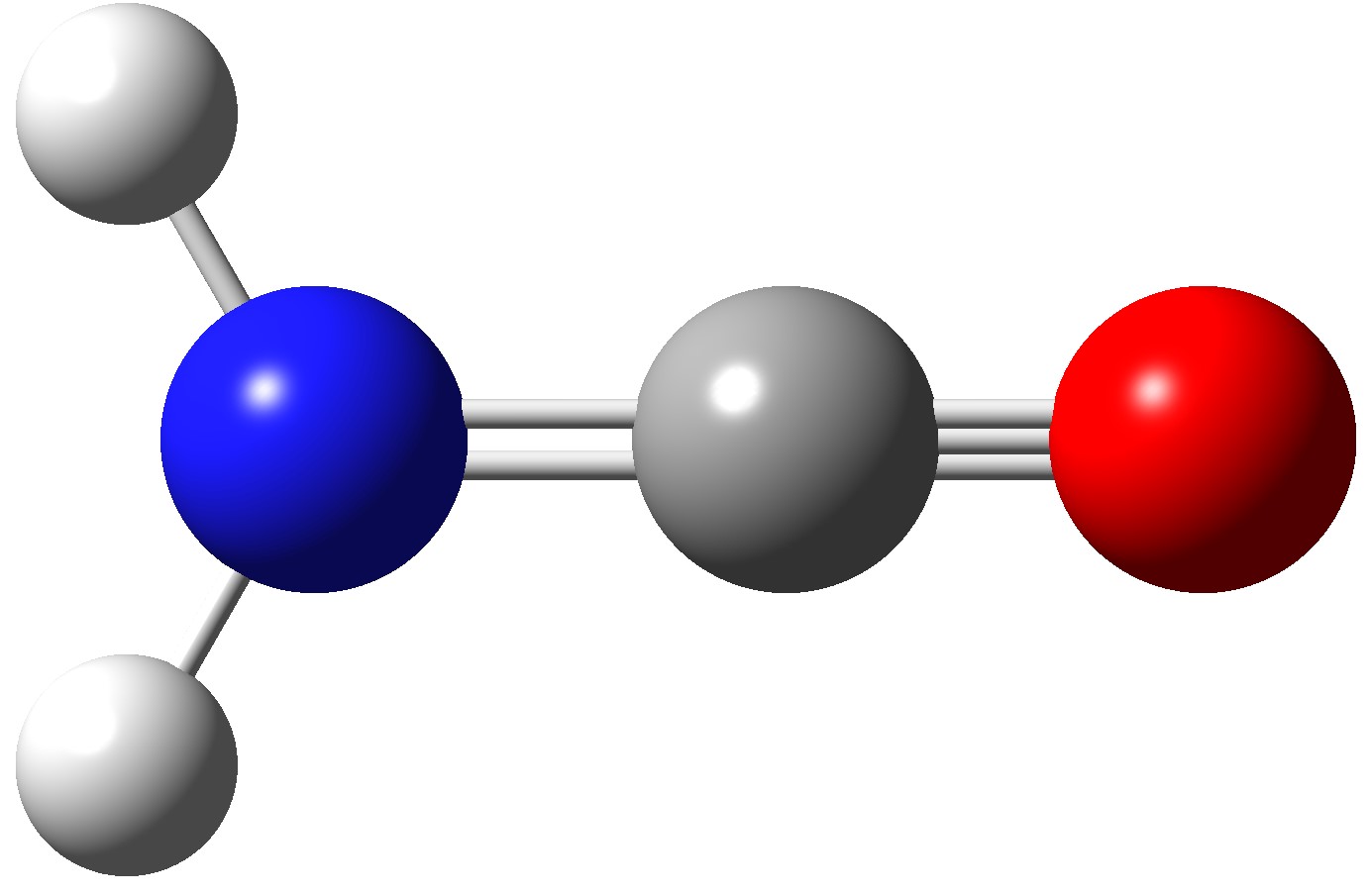}}   &        & 324210.9   & 2.9     & 3.8  \\
                                                                     &  0.0   & 10295.5    & 380.4   & 0.0  \\
        H$_{2}$NCO$^{+}$                                             &        & 9978.6     &         & 0.0  \\ \hline
        \multirow{2}{*}{\includegraphics[height=0.5cm]{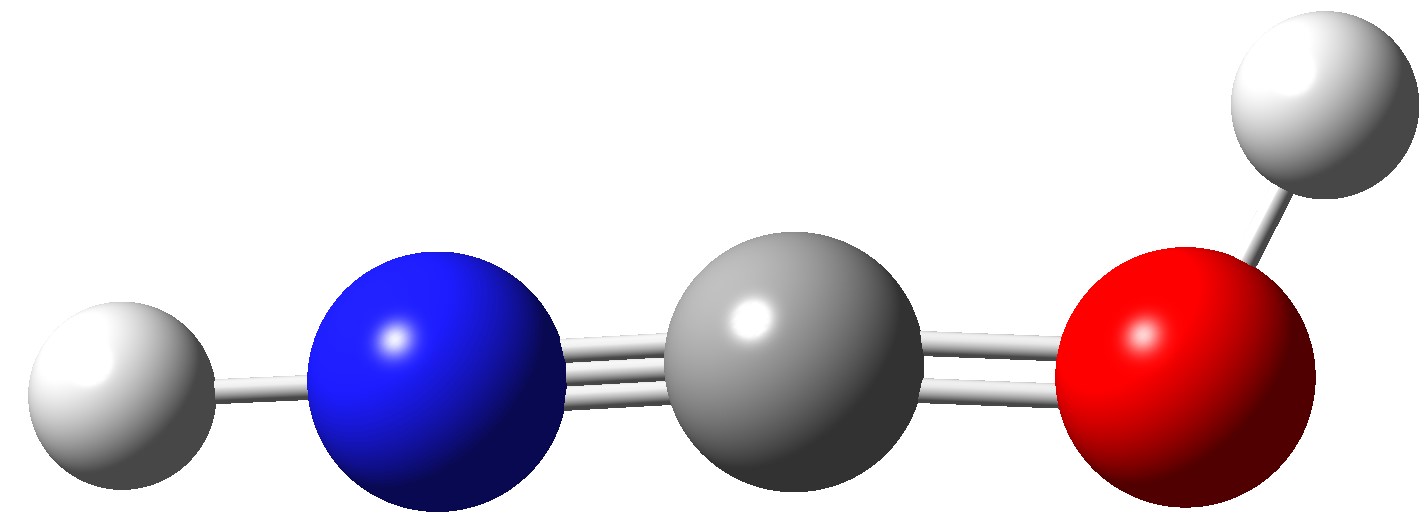}}   &        & 723374.4   & 2.8     & 1.3  \\
                                                                     &  70.4  & 10034.0   &1255.3    & 1.7  \\
        HNCOH$^{+}$                                                  &        & 9896.7    &          & 0.0  \\ \hline
        \multirow{2}{*}{\includegraphics[height=0.5cm]{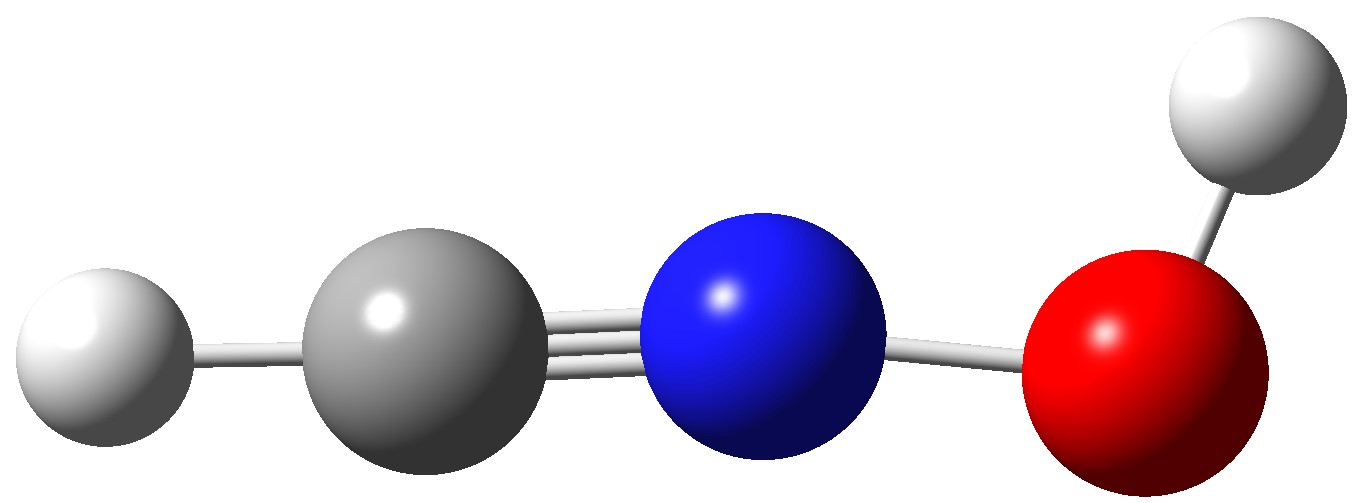}}   &        & 625942.1  &3.8       & 2.4  \\
                                                                     &  285.4 & 10400.0    &1859.1   & 1.6  \\
        HCNOH$^{+}$                                                  &        & 10230.0    &         & 0.0  \\ \hline
        \multirow{2}{*}{\includegraphics[height=0.5cm]{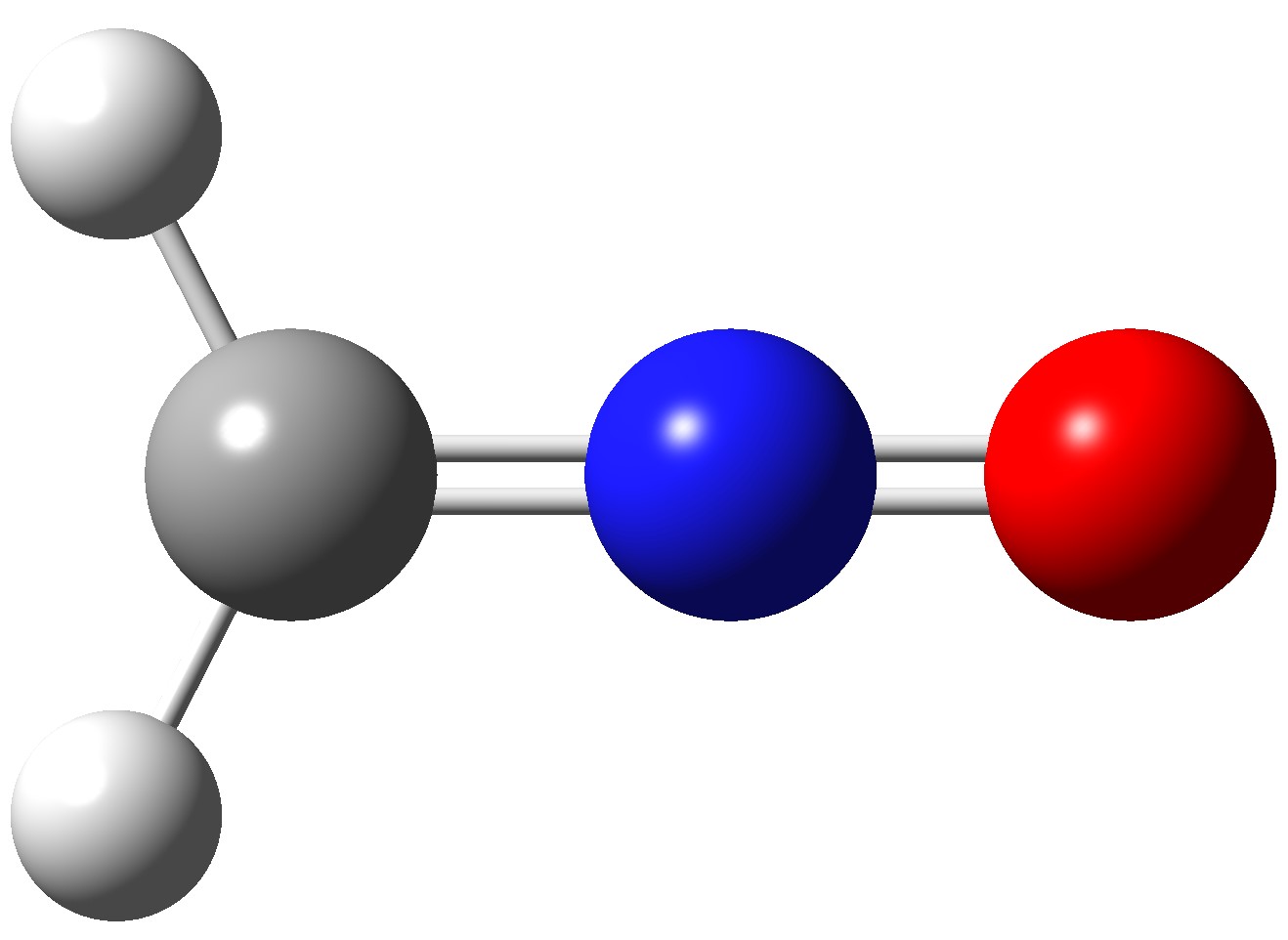}}   &        & 268002.3  & 3.6      & 2.9  \\
                                                                     &  333.8 & 11140.1   & 461.1    & 0.0  \\
        H$_{2}$CNO$^{+}$                                             &        & 10695.5  &           & 0.0  \\ \hline
        \multirow{2}{*}{\includegraphics[height=0.5cm]{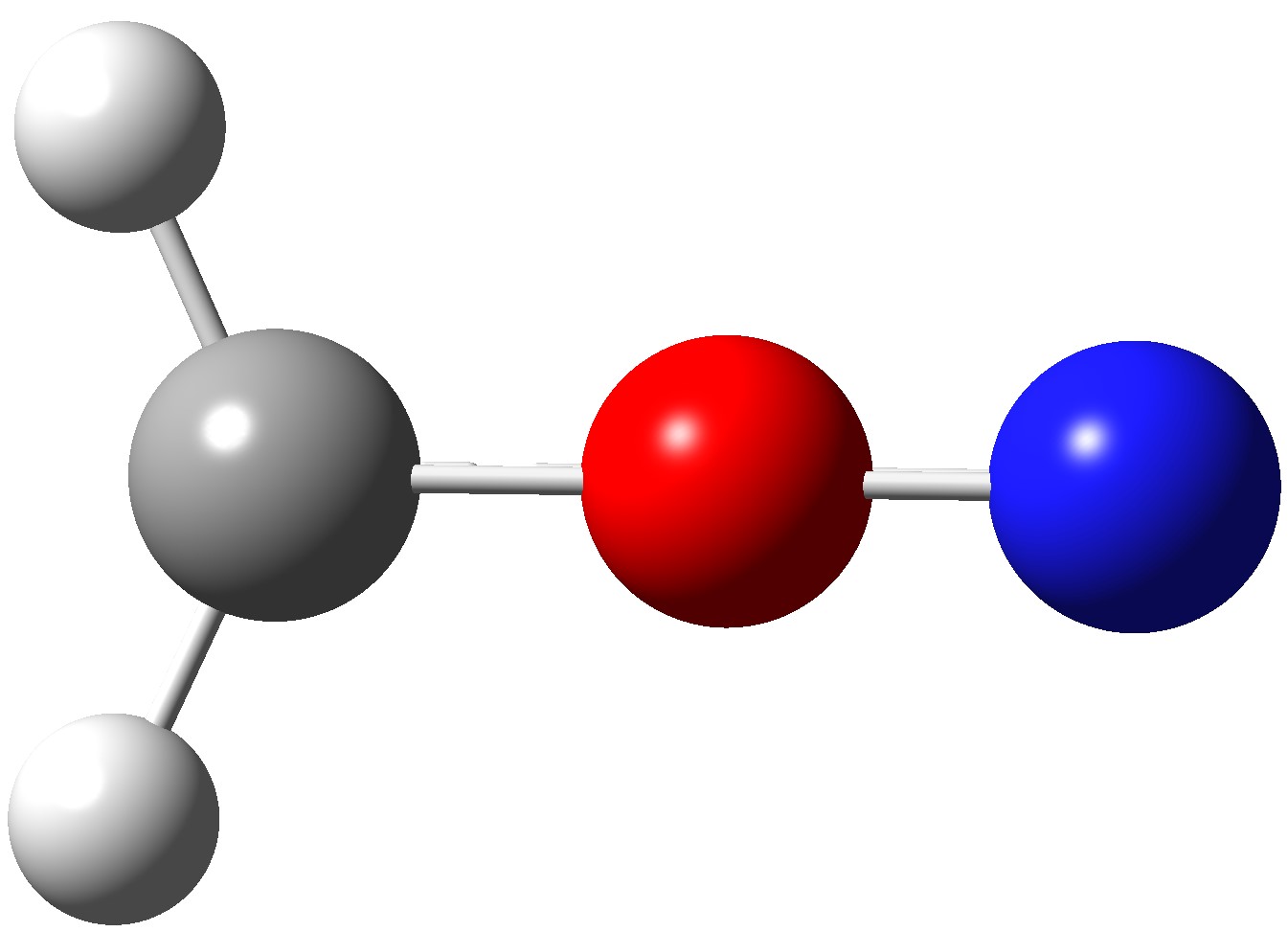}}   &        & 259994.7   & 4.6     & 2.3  \\
                                                                     & 613.2  & 11321.2   & 484.7    & 0.0  \\
        H$_{2}$CON$^{+}$                                             &        & 10848.8  &           & 0.0  \\ \hline\hline
\end{tabular}
\tablefoot{
        \tablefoottext{a}{Calculations at the CCSD/cc-pVTZ level of theory.} \tablefoottext{b}{The energy taken as reference is the lowest energy species within the same isomer family}.\\
}
\end{table*}

\begin{table*}
\caption{Scaled theoretical values for the spectroscopic parameters of CH$_3$CO$^+$, CH$_2$COH$^+$, and CH$_3$NCH$^+$ (all in MHz).}
\label{abini}
\centering
\begin{tabular}{{lccccc}}
\hline
\hline
&\multicolumn{2}{c}{CH$_3$CN}&\multicolumn{3}{c}{CH$_3$CO$^+$} \\
\cmidrule(lr){2-3} \cmidrule(lr){4-6}
Parameter & Calc.\tablefootmark{a} & Exp.\tablefootmark{b} & Calc.\tablefootmark{a} & Scaled\tablefootmark{c}
& Exp.\tablefootmark{d} \\
\hline
($A$-$B$)$\times10$$^{-3}$ &  150.4    &  148.900074(65)   &  145.0    &  143.5    &              -    \\
$B$                        &  9191.5   &  9198.899134(11)  &  9122.3   &  9129.6   &       9134.4742(8)\\
$D_J\times$10$^{3}$       &  3.6      &   3.807528(9)     &   3.7     &  3.9      &       4.014(13)   \\
$D_{JK}\times$10$^{3}$    &  174.0    &   177.40796(28)   &   181.0   &  184.6    &      188.47(50)   \\
\hline
\hline
&\multicolumn{2}{c}{CH$_2$CNH}&\multicolumn{3}{c}{CH$_2$COH$^+$} \\
\cmidrule(lr){2-3} \cmidrule(lr){4-6}
Parameter & Calc.\tablefootmark{a} & Exp.\tablefootmark{e} & Calc.\tablefootmark{a} & Scaled\tablefootmark{f}
& Exp.\tablefootmark{d} \\
\hline
$A$                     &  198393.1 &   201443.685(75)   &   198278.8    & 201327.7  &      -       \\
$B$                     &  9666.3   &   9663.168(2)      &   9411.8      & 9408.7   &      -       \\
$C$                     &  9488.0   &   9470.127(2)      &   9227.6      & 9210.2     & 9134.4742(8) \\
$D_J\times$10$^{3}$    &  2.9       &   2.980(2)         &   2.7         &  2.8      &  4.014(13)   \\
$D_{JK}\times$10$^{3}$ &  244.4     &   232.8(3)         &   397.7       &  378.7    & 188.47(50)   \\
\hline
\hline
&\multicolumn{2}{c}{CH$_3$NC}&\multicolumn{3}{c}{CH$_3$NCH$^+$} \\
\cmidrule(lr){2-3} \cmidrule(lr){4-6}
Parameter & Calc.\tablefootmark{a} & Exp.\tablefootmark{g} & Calc.\tablefootmark{a} & Scaled\tablefootmark{h}
& Exp.\tablefootmark{d} \\
\hline
($A$-$B$)$\times10$$^{-3}$ &  169.0    &  167.36100(23)   & 164.5     &   162.9    &          -      \\
$B$                        &  10049.0  &  10052.88568(25) & 9102.0    &   9105.5   &    9134.4742(8) \\
$D_J\times$10$^{3}$       &  4.4      &    4.69212(18)   & 3.8        &   4.0     &     4.014(13)   \\
$D_{JK}\times$10$^{3}$    &  222.9    &   227.5116(83)   & 168.2      &   171.7   &     188.47(50)  \\
\hline
\end{tabular}
\tablefoot{
\tablefoottext{a}{Rotational constants calculated at the CCSD(T)-F12b/aug-cc-pVQZ level of theory, and centrifugal distortion constants calculated
at the MP2/aug-cc-pVQZ level of theory.}
\tablefoottext{b}{\citet{Muller2009}.}
\tablefoottext{c}{Scaled by the ratio Exp/Calc. of the corresponding parameter for CH$_3$CN species}.
\tablefoottext{d}{This work.}
\tablefoottext{e}{\citet{Rodler1984}.}
\tablefoottext{f}{Scaled by the ratio Exp/Calc. of the corresponding parameter for CH$_2$CNH species}.
\tablefoottext{g}{\citet{Pliva1995}.}
\tablefoottext{h}{Scaled by the ratio Exp/Calc. of the corresponding parameter for CH$_3$NC species}.
}
\end{table*}

\section{Additional quantum chemical calculations for CH$_3$CO$^+$}
\label{more_calcu}
\begin{figure*}
\includegraphics[scale=0.33]{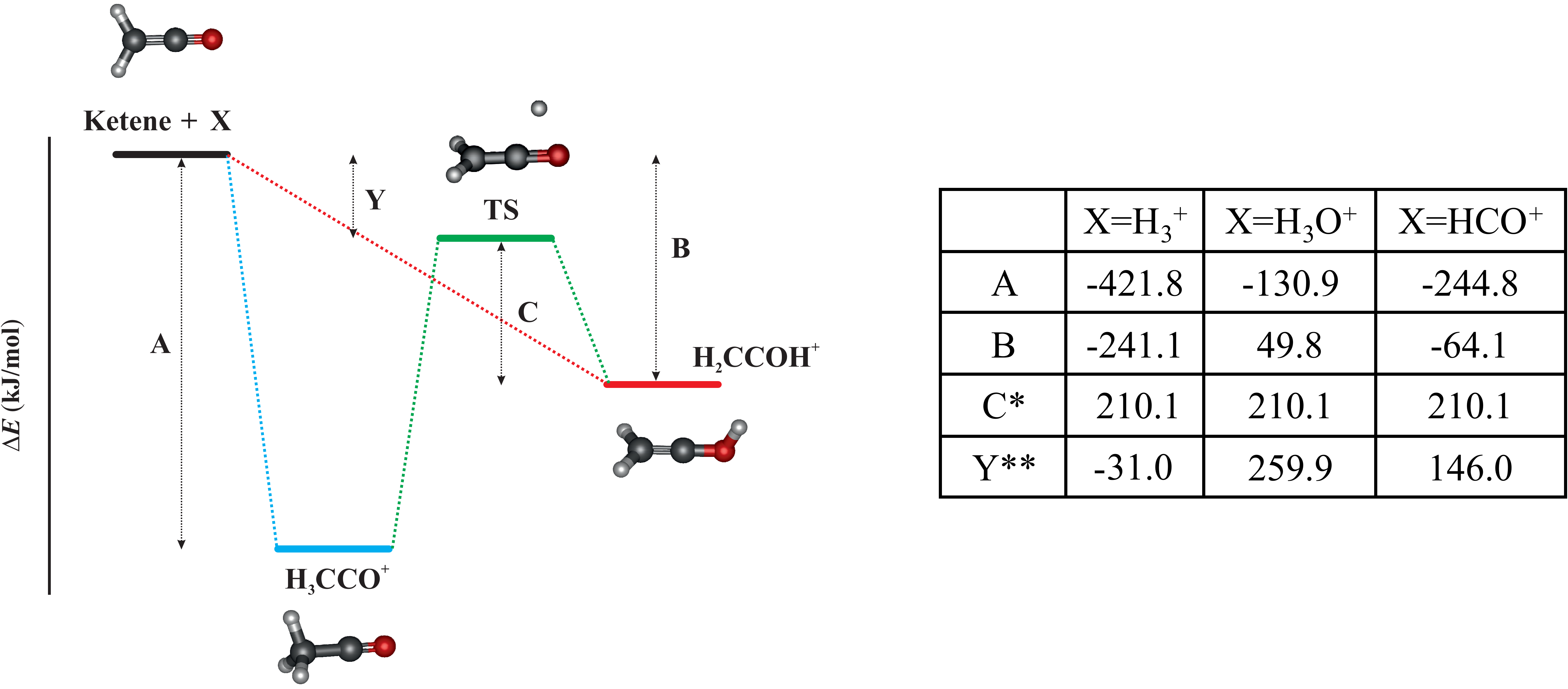}
\centering
\caption{\label{pes} Energy diagram for the protonation of ketene. Total energies relative to those of the separated reactants, ketene and the proton donor X, are given in the enclosed table in kJ mol$^{-1}$.
C* is the TS energy for the interconversion between CH$_2$COH$^+$ and CH$_3$CO$^+$
isomers. Y** is the energy difference between the reactants and the interconversion TS; a negative
value indicates that the TS is submerged below the reactant energy, and a positive value implies
that the TS lies above the reactant energy.}
\end{figure*}

The potential energy surface (PES) for the protonation of ketene has been explored at the CCSD/cc-pVTZ
level of theory. In the calculations, we considered three possible proton donors: H$_3^+$, H$_3$O$^+$,
and HCO$^+$, as well as the formation of the two isomers of protonated ketene, CH$_3$CO$^+$ and
CH$_2$COH$^+$. Figure \ref{pes} depicts the PES along the reaction coordinate for
the protonation of ketene and the relative energies for all the stationary points when ketene
reacts with H$_3^+$, H$_3$O$^+$, or HCO$^+$. For each reaction, the two reactants, ketene and
the proton donor, that separated from each other were assumed to be the energy zero.
The protonation of ketene in the CH$_2$ in the three cases is exothermic, and it proceeds
without any transition state (TS) to form CH$_3$CO$^+$. This formation is more favourable in the
case of H$_3^+$. On the other hand, the formation of CH$_2$COH$^+$ is exothermic in the cases
of H$_3^+$ and HCO$^+$, but endothermic in the case of H$_3$O$^+$. The less stable isomer,
CH$_2$COH$^+$, can interconvert through a hydrogen migration to CH$_3$CO$^+$, which has a
TS barrier height of 210.1 kJ/mol. As shown in Figure \ref{pes}, the TS
for this interconversion lies over the energy of the reactants in the protonation of ketene
with H$_3$O$^+$ and HCO$^+$. In contrast, this TS lies below the energy of
the reactants when ketene reacts with H$_3^+$.

\section{CH$_3$X species in TMC-1}
\label{CH3_X}
\begin{figure}
\includegraphics[scale=0.65]{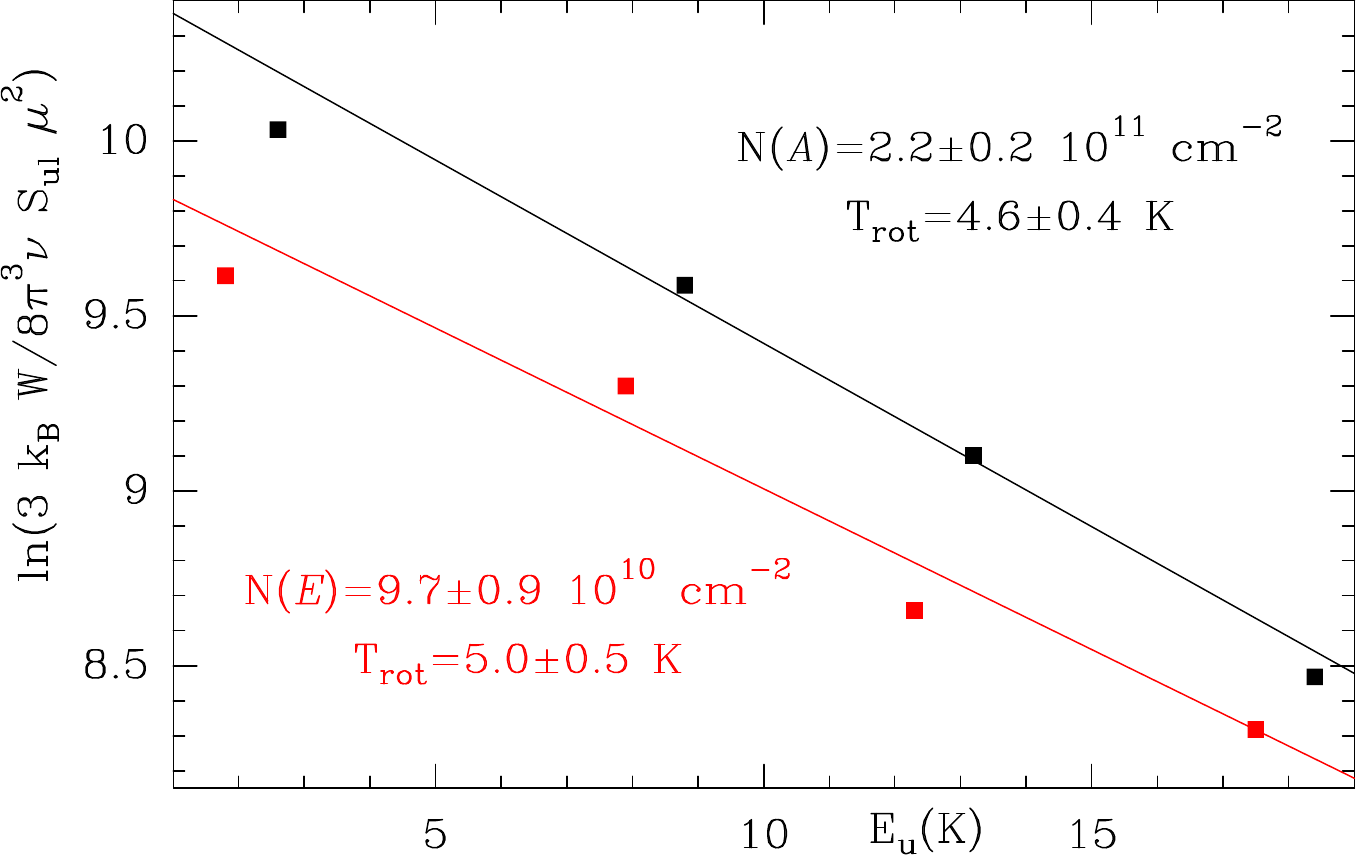}
\centering
\caption{\label{trot} Rotational diagrams for the $A$ (black line) and $E$ (red line) symmetry species of CH$_3$CO$^+$ in TMC-1.
}
\end{figure}

As noted above, the intensity of the $J$=2-1 $K$=1 line in TMC-1 is well below the expected
value if the rotational temperature for the $A$ and $E$ species is the same. In order to
check this point, we show in Fig. \ref{trot} the rotational diagrams for the $A$ and $E$ species of CH$_3$CO$^+$ using
the observed line parameters given in Table \ref{tab_CH3CO+}. The observed intensities have been corrected
for beam dilution and the beam efficiencies of the Yebes 40m and IRAM 30m telescopes.
We assumed a uniform source of radius 40$''$ \citep{Fosse2001}.
The derived rotational temperatures,
T$_{rot}(A)$=4.4$\pm$0.4 K and T$_{rot}(E)$=5.0$\pm$0.5 K, are consistent with a common excitation
through collisions with H$_2$. The derived column densities are N($A$-CH$_3$CO$^+$)=(2.2$\pm$0.2)$\times$10$^{11}$ cm$^{-2}$
and N($E$-CH$_3$CO$^+$)=(9.7$\pm$0.9)$\times$10$^{10}$ cm$^{-2}$. Hence, as discussed in Sect. \ref{interconversion}, 
the $A/E$ abundance
ratio for CH$_3$CO$^+$ has been modified through collisions with H$^+$, H$_3$$^+$, HCO$^+$, and H$_3$O$^+$. This
is a similar effect to that found in cold molecular clouds for molecules having ortho and para symmetry species.

\begin{figure}
\includegraphics[scale=0.65]{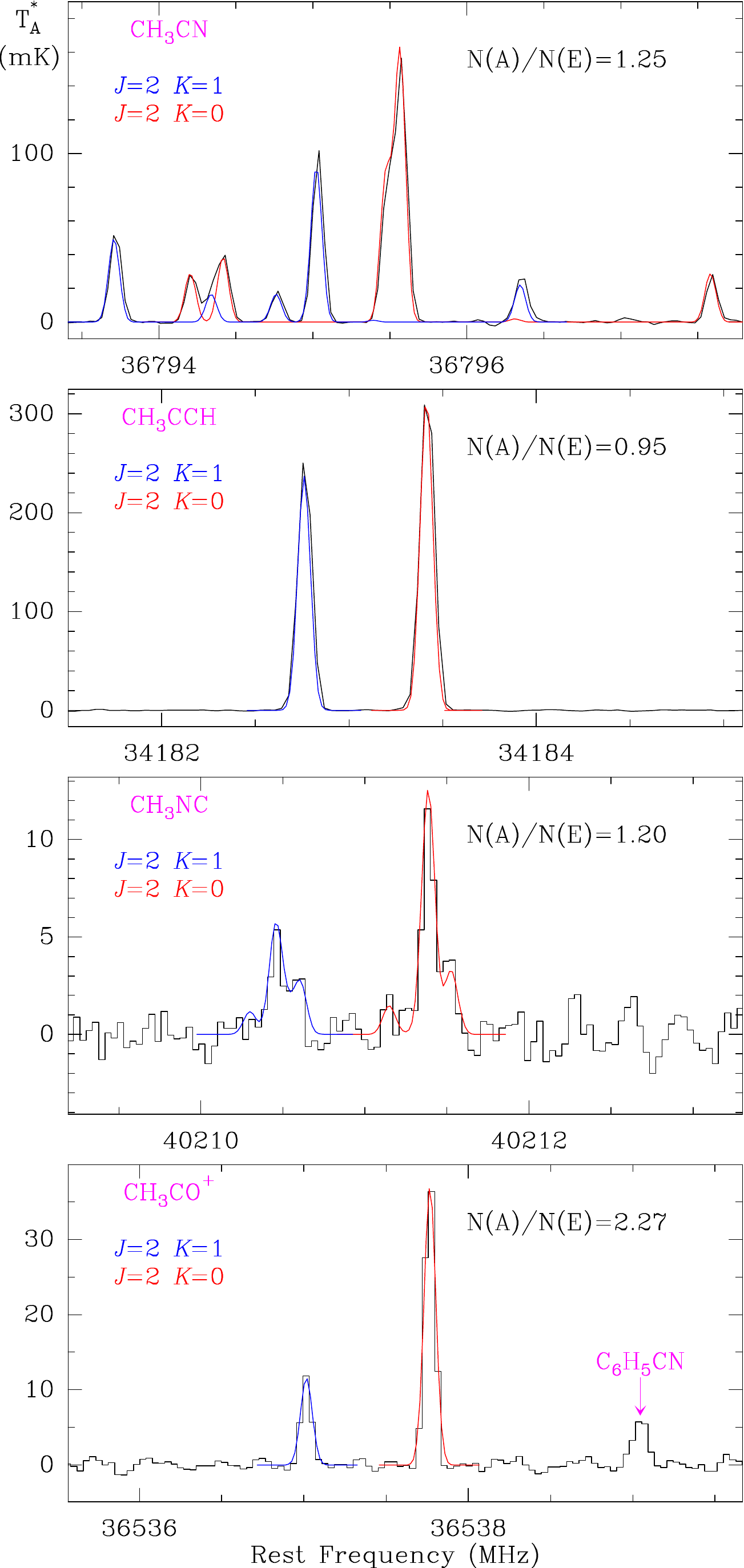}
\centering
\caption{\label{ae_rotor} Observed lines in the transition $J$=2-1 $K$=0,1
of different symmetric rotors in TMC-1. The colour lines represent the expected
line profiles for the $A$ species (red) and $E$ species (blue). The abundance ratio
between them in the model is indicated in each panel.}
\end{figure}

In order to check this peculiar result, we analysed all
symmetric rotors having transitions within our line survey: CH$_3$CCH,
CH$_3$C$_4$H,
CH$_3$CN,
and CH$_3$NC.
The cation CH$_3$CNH$^+$ has not been detected in TMC-1 (see Sect. \ref{dis_chem}).
The symmetric top CH$_3$C$_3$N is discussed in \cite{Marcelino2020b}.
Figure \ref{ae_rotor} shows the $J$=2-1 transition for CH$_3$CN, CH$_3$CCH, CH$_3$NC, and CH$_3$CO$^+$ (for this
species, see also Fig. \ref{fig_ch3co+}). The $K$=0 and $K$=1 lines of CH$_3$CN exhibit the
typical hyperfine structure introduced by the quadruple moment of the N nucleus. For all these additional
molecules, we assumed a rotational temperature of 10 K and a source radius of 40$''$ \citep{Fosse2001},
and we produced a synthetic spectrum that
is compared to the observations. We found that the $A/E$ abundance ratio is $\simeq$1 for all species but
CH$_3$CO$^+$. Adopting a lower rotational temperature has little effect on the derived $A/E$ abundance ratio
for these symmetric rotors.

\section{Observed and calculated frequencies of CH$_3$CO$^+$}
The frequencies observed in space and in the laboratory were merged
to obtain the recommended rotational and distortion constants. A total of
89 rotational transitions, ten in space (see Table \ref{tab_CH3CO+}) and 79 in the laboratory
(see Table \ref{freq_labo}), were fitted to the standard Hamiltonian of a
symmetric rotor \citep{Gordy1984}. For the lines observed in TMC-1 and other dark clouds, only
$B$, $D_J$, and $D_{JK}$ can be obtained as only rotational transitions with $K$=0 and 1
have been observed. The results are given in Table \ref{tab_fits}. For the 79 lines
observed in the laboratory, the constants $H_{JK}$ and H$_{KJ}$ were included in the
fit, and the results are given in Table \ref{tab_fits}. Finally, the merged fit to the
astronomical and laboratory lines produces the recommended set of rotational constants
given in the last column of Table \ref{tab_fits}. The observed and calculated frequencies,
together with the observed minus calculated values for the merged fit, are given in
Table \ref{freq_labo}.

We used the rotational and distortion constants that resulted
from the merged fit to the astronomical and laboratory lines
(see Table \ref{tab_fits}) to produce frequency predictions,
frequency uncertainties, line strengths, upper energy
levels, and Einstein coefficients for all transitions involving levels
with energies below 2000 K. They are given in Table \ref{freq_pred}. The
whole table is electronically available at the CDS. It should be noted that this table
contains the transitions for the $A$ and $E$ species, and that the $E$ lowest energy level,
$J_K$=$1_1$, is 7.8 K above the $0_0$ level of the $A$ species.

\clearpage
\onecolumn
\begin{longtable}{ccccccc}
\caption[]{Observed and calculated frequencies (in MHz) for CH$_3$CO$^+$.
\label{freq_labo}}\\
\hline
\hline
  J$_u$ & K   &  Freq. Obser.  &  (Unc)   &   Freq. Calc.  &  (Unc) & Obs-Calc   \\
\hline
\endfirsthead
\caption{continued.}\\
\hline
  J$_u$ &  K   &  Freq. Obser.  &  (Unc)   &   Freq. Calc.  &  (Unc) & Obs-Calc   \\
\hline
\endhead
\hline
\endfoot
\hline
\endlastfoot
\hline
     2& 0&   36537.765& 0.010&   36537.761 & 0.001&  0.004\\
     2& 1&   36537.014& 0.010&   36537.010 & 0.001&  0.004\\
     4& 0&   73074.769& 0.010&   73074.755 & 0.002&  0.014\\
     4& 1&   73073.252& 0.010&   73073.253 & 0.002& -0.001\\
     5& 0&   91342.732& 0.010&   91342.725 & 0.003&  0.007\\
     5& 1&   91340.865& 0.010&   91340.848 & 0.003&  0.017\\
     6& 0&  109610.225& 0.010&  109610.215 & 0.003&  0.010\\
     6& 1&  109607.954& 0.010&  109607.963 & 0.003& -0.009\\
     7& 0&  127877.133& 0.025&  127877.131 & 0.004&  0.002\\
     7& 1&  127874.494& 0.050&  127874.504 & 0.004& -0.010\\
    10& 3&  182639.714& 0.050&  182639.724 & 0.011& -0.010\\
    10& 2&  182658.451& 0.050&  182658.469 & 0.006& -0.018\\
    10& 1&  182669.687& 0.050&  182669.719 & 0.005& -0.032\\
    10& 0&  182673.442& 0.050&  182673.470 & 0.005& -0.028\\
    11& 3&  200900.043& 0.050&  200900.014 & 0.012&  0.029\\
    11& 2&  200920.617& 0.050&  200920.630 & 0.007& -0.013\\
    11& 1&  200933.010& 0.050&  200933.002 & 0.006&  0.008\\
    11& 0&  200937.148& 0.050&  200937.127 & 0.006&  0.021\\
    12& 3&  219159.236& 0.050&  219159.253 & 0.013& -0.017\\
    12& 2&  219181.747& 0.050&  219181.737 & 0.008&  0.010\\
    12& 1&  219195.209& 0.050&  219195.232 & 0.006& -0.023\\
    12& 0&  219199.712& 0.050&  219199.731 & 0.006& -0.019\\
    13& 3&  237417.332& 0.050&  237417.345 & 0.014& -0.013\\
    13& 2&  237441.706& 0.050&  237441.696 & 0.008&  0.010\\
    13& 1&  237456.320& 0.050&  237456.311 & 0.007&  0.009\\
    13& 0&  237461.178& 0.050&  237461.184 & 0.007& -0.006\\
    14& 3&  255674.171& 0.050&  255674.193 & 0.015& -0.022\\
    14& 2&  255700.388& 0.050&  255700.410 & 0.009& -0.022\\
    14& 1&  255716.107& 0.050&  255716.145 & 0.007& -0.038\\
    14& 0&  255721.370& 0.050&  255721.391 & 0.007& -0.021\\
    16& 3&  292183.778& 0.050&  292183.777 & 0.017&  0.001\\
    16& 2&  292213.646& 0.050&  292213.722 & 0.010& -0.075\\
    16& 1&  292231.636& 0.050&  292231.694 & 0.008& -0.058\\
    16& 0&  292237.659& 0.050&  292237.685 & 0.008& -0.026\\
    17& 3&  310436.291& 0.050&  310436.322 & 0.019& -0.031\\
    17& 2&  310468.201& 0.100&  310468.127 & 0.011&  0.074\\
    17& 1&  310487.210& 0.050&  310487.216 & 0.009& -0.006\\
    17& 0&  310493.523& 0.050&  310493.580 & 0.009& -0.057\\
    18& 3&  328687.226& 0.050&  328687.241 & 0.020& -0.015\\
    18& 2&  328720.866& 0.050&  328720.905 & 0.012& -0.039\\
    18& 1&  328741.056& 0.050&  328741.110 & 0.010& -0.054\\
    18& 0&  328747.827& 0.050&  328747.846 & 0.010& -0.019\\
    21& 6&  383217.758& 0.050&  383217.721 & 0.096&  0.037\\
    21& 5&  383303.931& 0.050&  383303.862 & 0.064&  0.069\\
    21& 4&  383374.395& 0.050&  383374.396 & 0.041& -0.001\\
    21& 3&  383429.306& 0.050&  383429.289 & 0.025&  0.017\\
    21& 2&  383468.517& 0.050&  383468.517 & 0.015&  0.000\\
    21& 1&  383492.044& 0.050&  383492.060 & 0.012& -0.016\\
    21& 0&  383499.892& 0.050&  383499.910 & 0.012& -0.018\\
    22& 6&  401451.192& 0.050&  401451.205 & 0.103& -0.013\\
    22& 5&  401541.411& 0.050&  401541.408 & 0.069&  0.003\\
    22& 4&  401615.264& 0.050&  401615.268 & 0.044& -0.004\\
    22& 3&  401672.742& 0.050&  401672.750 & 0.027& -0.008\\
    22& 2&  401713.849& 0.050&  401713.827 & 0.017&  0.022\\
    22& 1&  401738.482& 0.050&  401738.481 & 0.014&  0.001\\
    22& 0&  401746.707& 0.050&  401746.700 & 0.014&  0.007\\
    23& 6&  419682.601& 0.050&  419682.598 & 0.111&  0.003\\
    23& 5&  419776.829& 0.050&  419776.858 & 0.074& -0.029\\
    23& 4&  419854.065& 0.050&  419854.039 & 0.048&  0.026\\
    23& 3&  419914.100& 0.050&  419914.106 & 0.029& -0.006\\
    23& 2&  419957.037& 0.050&  419957.031 & 0.018&  0.006\\
    23& 1&  419982.794& 0.050&  419982.794 & 0.015&  0.000\\
    23& 0&  419991.387& 0.050&  419991.383 & 0.015&  0.004\\
    24& 6&  437911.779& 0.050&  437911.807 & 0.119& -0.026\\
    24& 5&  438010.092& 0.050&  438010.117 & 0.080& -0.025\\
    24& 4&  438090.613& 0.050&  438090.615 & 0.052& -0.002\\
    24& 3&  438153.261& 0.050&  438153.263 & 0.032& -0.002\\
    24& 2&  438198.039& 0.050&  438198.033 & 0.020&  0.006\\
    24& 1&  438224.913& 0.050&  438224.903 & 0.016&  0.010\\
    24& 0&  438233.877& 0.050&  438233.861 & 0.016&  0.016\\
    25& 6&  456138.764& 0.150&  456138.735 & 0.129&  0.029\\
    25& 5&  456240.963& 0.150&  456241.090 & 0.087& -0.127\\
    25& 4&  456324.889& 0.050&  456324.900 & 0.056& -0.011\\
    25& 3&  456390.167& 0.050&  456390.126 & 0.034&  0.041\\
    25& 2&  456436.758& 0.050&  456436.737 & 0.022&  0.021\\
    25& 1&  456464.703& 0.050&  456464.712 & 0.018& -0.009\\
    25& 0&  456474.048& 0.050&  456474.039 & 0.018&  0.009\\
    26& 5&  474469.660& 0.050&  474469.681 & 0.094& -0.021\\
    26& 4&  474556.835& 0.050&  474556.798 & 0.061&  0.037\\
    26& 3&  474624.604& 0.050&  474624.597 & 0.037&  0.007\\
    26& 2&  474673.034& 0.050&  474673.048 & 0.024& -0.014\\
    26& 1&  474702.158& 0.050&  474702.127 & 0.020&  0.031\\
    26& 0&  474711.811& 0.050&  474711.821 & 0.020& -0.010\\
    27& 5&  492695.931& 0.150&  492695.796 & 0.102&  0.135\\
    27& 4&  492786.129& 0.150&  492786.214 & 0.066& -0.085\\
    27& 3&  492856.584& 0.050&  492856.582 & 0.041&  0.002\\
    27& 2&  492906.862& 0.050&  492906.869 & 0.026& -0.007\\
    27& 1&  492937.060& 0.050&  492937.050 & 0.022&  0.010\\
    27& 0&  492947.117& 0.050&  492947.112 & 0.022&  0.005\\
\hline
\end{longtable}
\twocolumn

\begin{table*}
\caption{Frequency predictions for CH$_3$CO$^+$ $^*$.}
\label{freq_pred}
\centering
\begin{tabular}{{lccccccccc}}
\hline
\hline
$J_u$$^a$ & $K_u$$^a$ & $J_l$$^a$ & $K_l$$^a$& $\nu$(MHz)$^b$ & Unc(MHz)$^c$ & E$_{up}$(K)$^d$& $A_{ij}$(s$^{-1}$)$^e$ & $S_{ij}$$^f$ & $g_u$$^g$\\
\hline
1 &  0 &  0 &  0 &   18268.92826&   0.00039&     0.9& 2.898$\times$10$^{-07}$&    1.0000&   3\\
2 &  1 &  1 &  1 &   36537.00979&   0.00072&     9.5& 2.086$\times$10$^{-06}$&    1.5000&   5\\
2 &  0 &  1 &  0 &   36537.76068&   0.00079&     2.6& 2.782$\times$10$^{-06}$&    2.0000&   5\\
3 &  2 &  2 &  2 &   54801.89688&   0.00115&    32.8& 5.587$\times$10$^{-06}$&    1.6667&   7\\
3 &  1 &  2 &  1 &   54805.27517&   0.00107&    12.1& 8.941$\times$10$^{-06}$&    2.6667&   7\\
3 &  0 &  2 &  0 &   54806.40143&   0.00117&     5.3& 1.006$\times$10$^{-05}$&    3.0000&   7\\
4 &  3 &  3 &  3 &   73061.24457&   0.00265&    70.7& 1.081$\times$10$^{-05}$&    3.5000&  18\\
4 &  2 &  3 &  2 &   73068.74904&   0.00151&    36.3& 1.854$\times$10$^{-05}$&    3.0000&   9\\
4 &  1 &  3 &  1 &   73073.25310&   0.00141&    15.7& 2.318$\times$10$^{-05}$&    3.7500&   9\\
4 &  0 &  3 &  0 &   73074.75469&   0.00154&     8.8& 2.473$\times$10$^{-05}$&    4.0000&   9\\
5 &  4 &  4 &  4 &   91312.71293&   0.00563&   123.3& 1.776$\times$10$^{-05}$&    1.8000&  11\\
5 &  3 &  4 &  3 &   91325.83851&   0.00326&    75.1& 3.159$\times$10$^{-05}$&    6.4000&  22\\
5 &  2 &  4 &  2 &   91335.21824&   0.00186&    40.7& 4.148$\times$10$^{-05}$&    4.2000&  11\\
5 &  1 &  4 &  1 &   91340.84780&   0.00173&    20.0& 4.741$\times$10$^{-05}$&    4.8000&  11\\
5 &  0 &  4 &  0 &   91342.72461&   0.00189&    13.2& 4.939$\times$10$^{-05}$&    5.0000&  11\\
6 &  5 &  5 &  5 &  109553.96919&   0.01021&   190.5& 2.644$\times$10$^{-05}$&    1.8333&  13\\
6 &  4 &  5 &  4 &  109574.20539&   0.00662&   128.6& 4.810$\times$10$^{-05}$&    3.3333&  13\\
6 &  3 &  5 &  3 &  109589.95432&   0.00384&    80.4& 6.496$\times$10$^{-05}$&    9.0000&  26\\
6 &  2 &  5 &  2 &  109601.20873&   0.00219&    46.0& 7.701$\times$10$^{-05}$&    5.3333&  13\\
6 &  1 &  5 &  1 &  109607.96344&   0.00203&    25.3& 8.425$\times$10$^{-05}$&    5.8333&  13\\
6 &  0 &  5 &  0 &  109610.21536&   0.00223&    18.4& 8.666$\times$10$^{-05}$&    6.0000&  13\\
7 &  6 &  6 &  6 &  127782.68995&   0.01698&   272.3& 3.683$\times$10$^{-05}$&    3.7143&  30\\
7 &  5 &  6 &  5 &  127811.51930&   0.01162&   196.6& 6.805$\times$10$^{-05}$&    3.4286&  15\\
7 &  4 &  6 &  4 &  127835.12506&   0.00755&   134.7& 9.361$\times$10$^{-05}$&    4.7143&  15\\
7 &  3 &  6 &  3 &  127853.49637&   0.00439&   86.5& 1.135$\times$10$^{-04}$&   11.4286&  30\\
7 &  2 &  6 &  2 &  127866.62477&   0.00249&    52.1& 1.278$\times$10$^{-04}$&    6.4286&  15\\
7 &  1 &  6 &  1 &  127874.50422&   0.00232&   31.4& 1.363$\times$10$^{-04}$&    6.8571&  15\\
7 &  0 &  6 &  0 &  127877.13111&   0.00254&    24.5& 1.391$\times$10$^{-04}$&    7.0000&  15\\
\hline
\end{tabular}
\tablefoot{\\
\tablefoottext{*}{The entire table is electronically available at the CDS
via anonymous ftp to cdsarc.u-strasbg.fr (130.79.128.5)
or via http://cdsweb.u-strasbg.fr/cgi-bin/qcat?J/A+A
}\\
\tablefoottext{a}{Upper and lower $J_K$ quantum numbers.}\\
\tablefoottext{b}{Predicted frequency (in MHz).}\\
\tablefoottext{c}{Uncertainty in the predicted frequency (in MHz).}\\
\tablefoottext{d}{Energy (in K) of the upper energy level of the transition.}\\
\tablefoottext{e}{Einstein coefficient of the transition (in s$^{-1}$).}\\
\tablefoottext{f}{Line strength.}\\
\tablefoottext{g}{Degeneracy of the upper level. It is 2$J$+1 for all levels
except for those with $K$=3 $\times$ n (n=1,2,...), for which $g_u$ is 2$\times$(2$J$+1).}\\
}
\end{table*}

\end{appendix}

\end{document}